\pdfoutput=1

\documentclass[ amsmath, amssymb, aps, prd, twocolumn, lengthcheck, sort&compress, showpacs, groupedaddress, nofootinbib ]{revtex4}

\usepackage{amsmath}
\usepackage{amssymb}
\usepackage{graphicx}
\usepackage{dsfont}
\usepackage{dcolumn}
\usepackage{units}
\usepackage{bm}

\usepackage[usenames]{color}

\usepackage[colorlinks=true,linkcolor=blue,citecolor=blue,urlcolor=blue]{hyperref}

      \newcommand{\conjg}[1]{\ensuremath{\hspace{1pt}\overline{\hspace{-1pt}#1\hspace{-1pt}}}\hspace{1pt}}
      \newcommand{\vect}[1]{\bm{#1}}

      \newcommand{\grad}{\vect{\nabla}}

\def\Slash#1{\setbox0=\hbox{$#1$} 
\dimen0=\wd0 
\setbox1=\hbox{/} \dimen1=\wd1 
\ifdim\dimen0>\dimen1 
\rlap{\hbox to \dimen0{\hfil/\hfil}} 
#1 
\else 
\rlap{\hbox to \dimen1{\hfil$#1$\hfil}} 
/ 
\fi}

\def\longlonglongrightarrow{
\relbar\joinrel\relbar\joinrel\relbar\joinrel\relbar\joinrel\relbar\joinrel\relbar\joinrel\rightarrow}
\def\longlonglonglongrightarrow{
\relbar\joinrel\relbar\joinrel\relbar\joinrel\relbar\joinrel\relbar\joinrel\relbar\joinrel\relbar\joinrel\relbar\joinrel\relbar\joinrel\relbar\joinrel\relbar\joinrel\rightarrow}

\begin{document}

         \title{Nucleon electromagnetic form factors\\
         from the covariant Faddeev equation}

         \author{G.~Eichmann}
         \affiliation{Institut f\"{u}r Theoretische Physik I, Justus-Liebig-Universit\"at Giessen, D-35392 Giessen, Germany  }
         \email{gernot.eichmann@theo.physik.uni-giessen.de}

         \date{\today}

         \begin{abstract}

             We compute the electromagnetic form factors of the nucleon in the Poincar\'e-covariant Faddeev framework
             based on the Dyson-Schwinger equations of QCD.
             The general expression for a baryon's electromagnetic current in terms of three interacting dressed quarks is derived.
             Upon employing a rainbow-ladder gluon-exchange kernel for the quark-quark interaction,
             the nucleon's Faddeev amplitude and electromagnetic form factors are computed without any further truncations or model assumptions.
             The form factor results show clear evidence of missing pion-cloud effects below a photon momentum transfer of $\sim 2$ GeV$^2$
             and in the chiral region
             whereas they agree well with experimental data at higher photon momenta.
             Thus, the approach reflects the properties of the nucleon's quark core.

         \end{abstract}


         \maketitle

        \begin{figure*}[tp]
                    \begin{center}

                    \includegraphics[scale=0.12]{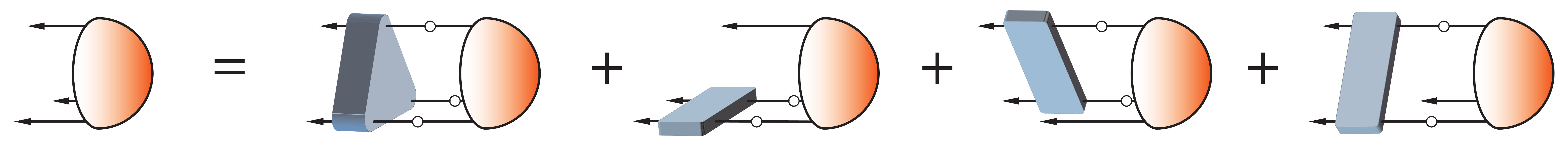}
                    \caption{(Color online) The covariant three-quark equation for a baryon amplitude, Eq.\,(\ref{three-quark-eq}).}\label{fig1}

                    \end{center}
        \end{figure*}


\section{Introduction}

        The measurement of the nucleon's elastic electromagnetic form factors has been subject to
        major experimental progress during the past years; see Refs.~\cite{Perdrisat:2006hj,Arrington:2006zm,Arrington:2011kb} for recent reviews.
        Precise polarization experiments at Jefferson Lab, MIT-Bates and MAMI have charted  
        the evolution of the electromagnetic form factors over a wide range of photon momenta,  
        and the forthcoming $12$-GeV upgrade at JLAB will allow for an extension of these measurements even farther beyond.  
        In connection to this, 
        the $N^\star$ program at JLAB is expected to provide substantial insight in the internal structure of nucleon resonances~\cite{Aznauryan:2009da,Aznauryan:2011ub}.

        As electromagnetic form factors encode the nucleon's spatial charge and magnetization distribution,
        they provide a direct connection to the underlying dynamics of quarks and gluons 
        which are the fundamental degrees of freedom in Quantum Chromodynamics (QCD).
        The new data have raised many new questions and stimulated a parallel response from theory.
        Among the aspects that have been intensely discussed 
        are the question of two-photon contributions to elastic cross sections~\cite{Guichon:2003qm,Arrington:2007ux,Carlson:2007sp};
        the importance of quark orbital angular momentum in the nucleon's structure at high momentum \cite{Belitsky:2002kj,Ralston:2003mt,Brodsky:2003pw,Thomas:2008ga}; 
        or the role of pseudoscalar meson clouds and
        the interpretation of form factors through charge and magnetization densities in different frames~\cite{Friedrich:2003iz,Meissner:2007tp,Miller:2008jc,Miller:2007kt,Crawford:2010gv}. 
        At the same time, advances in studying hard exclusive processes
        have provided insight into the spatial and spin structure of the nucleon 
        through the framework of generalized parton distributions~\cite{Goeke:2001tz,Diehl:2003ny,Belitsky:2005qn,Burkardt:2008jw}.

       Approaching these issues from a fundamental theoretical perspective
       aims at a quantitative understanding of the hadrons' substructure in terms of quarks and gluons in QCD.
       A comprehensive theoretical framework which has seen notable progress in recent years  
       is given by the Dyson-Schwinger equations (DSEs) of QCD.
       They form a set of infinitely many coupled integral equations for QCD's Green functions
       that grants access to nonperturbative phenomena such as dynamical chiral symmetry breaking, confinement, and the formation of bound states;
       see~\cite{Roberts:1994dr,Alkofer:2000wg,Fischer:2006ub} for reviews.

       The bridge between Dyson-Schwinger equations and hadron properties is provided by covariant bound-state equations,
       see~\cite{Maris:2003vk,Roberts:2007jh,Eichmann:2009zx} and references therein.
       In the case of baryons, it is the covariant Faddeev equation~\cite{Faddeev:1960su,Taylor:1966zza,Boehm:1976ya,Loring:2001kv}
       that describes a baryon's amplitude through all interactions that can take place between its dressed valence quarks,
       and thereby constitutes the direct analogue to the Bethe-Salpeter equation for mesons~\cite{Salpeter:1951sz}.
       The Faddeev equation was recently solved for the nucleon mass~\cite{Eichmann:2009qa,Eichmann:2009en}
       with an interaction kernel that traces the nucleon's binding to its quark-quark correlations and
       thereby relates the properties of baryons to those of mesons and the underlying substructure in QCD.
       First steps toward the $\Delta-$baryon in that setup were reported in Ref.~\cite{SanchisAlepuz:2010in},
       and the heavy-quark limit of the Faddeev equation was recently studied in Coulomb gauge~\cite{Popovici:2010ph}.

        In order to compute electromagnetic form factors one must additionally specify how the photon couples to the dressed quarks in a baryon.
        Due to the complexity of the three-quark problem, baryon form-factor calculations
        within the Dyson-Schwinger framework have been performed in the covariant quark-diquark model~\cite{Burden:1988dt,Ishii:1995bu,Hellstern:1997pg,Oettel:1998bk,Bloch:1999ke,Oettel:1999gc,Oettel:2000jj,Oettel:2002wf,Bloch:2003vn,Alkofer:2004yf,Cloet:2008re}    
        which traces the nucleon's binding and its interaction with the photon
        to scalar and axial-vector diquark correlations in its interior.
        In Refs.~\cite{Eichmann:2007nn,Eichmann:2008ef} the quark-diquark model was extended insofar as the dynamics of the quarks and diquarks
        were determined from their quark and gluon constituents,
        and corresponding results for nucleon and $\Delta-$baryon form factors were reported in Refs.~\cite{Eichmann:2009zx,Eichmann:2010je,Nicmorus:2010sd}.
        These studies provide valuable insight in the correlations that take place amongst the nucleon's constituents when probed by a photon,
        and several characteristic features revealed in experiments are readily reproduced.
        Nevertheless, the predictive power of the quark-diquark approach is limited by the sensitivity to the diquark parameters at larger photon momentum.  

        In the present work we follow Ref.~\cite{Eichmann:2009qa} and solve the covariant Faddeev equation directly without introducing explicit diquark degrees of freedom.
        The nucleon's binding is described by a rainbow-ladder (RL) interaction which amounts to an iterated dressed gluon exchange between any two quarks.
        The single parameter of the approach is a scale which is fixed to meson properties.
        We construct a consistent electromagnetic current operator and compute the nucleon's electromagnetic form factors in this setup.
        A rainbow-ladder truncation excludes the presence of meson-cloud corrections which have significant impact on
        the chiral and low-momentum structure of hadrons, hence the approach reflects the properties of a hadronic quark core.

        The manuscript is organized as follows: in Section~\ref{sec:Faddeev} we recall the Poincar\'{e}-covariant
        Faddeev equation, collect its ingredients, and discuss the nucleon amplitude and the current-mass evolution of its mass.
        In Section~\ref{sec:ffs} we construct the nucleon's electromagnetic current operator in the three-quark framework.
        Results for the nucleon's electromagnetic form factors are presented in Section~\ref{sec:results}.
        Technical details of the calculation are relegated to the appendices:
        conventions and formulas are collected in App.~\ref{app:conventions};
        the structure of the nucleon amplitude, its permutation-group properties, basis decomposition and partial-wave analysis are analyized in App.~\ref{sec:nucamp};
        App.~\ref{sec:faddeeveqsolution} presents an updated solution strategy for the Faddeev equation;
        and the nucleon's electromagnetic current operator and its ingredients in the three-quark framework are described in App.~\ref{sec:current}.
        Throughout this paper we work in Euclidean momentum space and use the isospin-symmetric limit $m_u=m_d$.


\section{Covariant Faddeev-equation} \label{sec:Faddeev}

            Baryons in QCD appear as poles in the three-quark scattering matrix.
            The residue at a particular pole defines the baryon's bound-state amplitude.
            It satisfies a covariant homogeneous integral equation  
            which is illustrated in Fig.~\ref{fig1} and reads
            \begin{equation}\label{three-quark-eq}
                \Psi = \widetilde{K}\,\Psi\,, \qquad \widetilde{K} = \widetilde{K}_\text{irr} + \sum_{a=1}^3 \widetilde{K}^{(a)}\,,
            \end{equation}
            where $\Psi$ is the bound-state amplitude defined on the baryon mass shell.  
            The three-body kernel $\widetilde{K}$
            comprises a three-quark irreducible contribution and the sum of permuted two-quark kernels
            whose quark-antiquark analogues appear in a meson BSE, and
            the subscript $a$ denotes the respective spectator quark.

            It has been a longstanding suggestion  
            that the strong attraction between two quarks in a $SU(3)_C$ antitriplet configuration within a baryon
            is a key feature for a better understanding of hadron properties~\cite{Anselmino:1992vg,Jaffe:2004ph}.
            This has been the guiding idea in the quark-diquark approach, and  
            in our context it motivates the omission of the three-body irreducible contribution,
            i.e. the first diagram on the right-hand side of Fig.~\ref{fig1},
            from the full three-quark kernel.
            The resulting covariant Faddeev equation
            attributes the binding mechanism of three quarks in a baryon to its quark-quark correlations and reads:
                \begin{equation}\label{faddeev:eq}
                \begin{split}
                    \Psi_{\alpha\beta\gamma\delta}(p, q,& \,P)  = \\ =\int\limits_k  \Big[ \, & \widetilde{K}_{\beta\beta'\gamma\gamma'}^{(1)} \, \Psi_{\alpha\beta'\gamma'\delta}(p^{(1)},q^{(1)},P) \,+   \\[-3.5mm]
                                                                                     & \widetilde{K}_{\gamma\gamma'\alpha\alpha'}^{(2)} \, \Psi_{\alpha'\beta\gamma'\delta}(p^{(2)},q^{(2)},P) \, + \\
                                                                                     & \widetilde{K}_{\alpha\alpha'\beta\beta'}^{(3)} \, \Psi_{\alpha'\beta'\gamma\delta}(p^{(3)},q^{(3)},P) \,\Big] \,,
                \end{split}
                \end{equation}
            where $\widetilde{K}^{(a)}$ denotes the renormalization-group invariant product of a $qq$ kernel and two dressed quark propagators:
            \begin{equation} \label{KSS}
                \widetilde{K}_{\alpha\alpha'\beta\beta'}^{(a)} = \mathcal{K}_{\alpha\alpha''\beta\beta''}\, S_{\alpha''\alpha'}(p_b-k)  \, S_{\beta''\beta'}(p_c+k) \,,
            \end{equation}
            and $\{a,b,c\}$ is an even permutation of $\{1,2,3\}$.

    \renewcommand{\arraystretch}{1.3}

            The notation in Eqs.~(\ref{faddeev:eq}--\ref{KSS}) is understood as follows:
            the Poincar\'e-covariant nucleon amplitude
            carries three spinor indices $\{\alpha,\beta,\gamma\}$ for the valence quarks
            and one index $\delta$ for the spin-$1/2$ nucleon.
            It depends on three quark momenta $p_1$, $p_2$, $p_3$ which can be expressed in terms of
            the total nucleon momentum $P$, where $P^2 = -M^2$ is fixed, and two relative Jacobi momenta $p$ and $q$.
            Upon choosing symmetric momentum partitioning
            they are related via:
                \begin{equation}
                \begin{array}{l}  p = \dfrac{2\,p_3-p_1-p_2}{3}\,,  \\[0.2cm]
                                  q = \dfrac{p_2-p_1}{2}\,,  \\[0.2cm]
                                  P = p_1+p_2+p_3\,,
                \end{array} \qquad
                \begin{array}{l}  p_1 =  -q -\dfrac{p}{2} + \dfrac{P}{3}\,, \\[0.2cm]
                                  p_2 =  q -\dfrac{p}{2} + \dfrac{P}{3}\,, \\[0.2cm]
                                  p_3 =  p + \dfrac{P}{3}\,.
                \end{array}
                \end{equation}
            The quark propagators $S$ depend on the internal quark momenta
            $p_i-k$ and $p_j+k$, where $k$ is the loop momentum that will later be identified with the gluon momentum, and
            the internal relative momenta in Eq.~\eqref{faddeev:eq} are given by:
    \renewcommand{\arraystretch}{1.2}
                \begin{equation}\label{fe-momenta-internal}
                \begin{array}{l}  p^{(1)} = p+k\,, \\
                                  p^{(2)} = p-k\,, \\
                                  p^{(3)} = p\,,
                \end{array} \qquad\quad
                \begin{array}{l}  q^{(1)} = q-k/2\,, \\
                                  q^{(2)} = q-k/2\,, \\
                                  q^{(3)} = q+k\,.
                \end{array}
                \end{equation}
         The inherent color structure leads to a prefactor $\nicefrac{2}{3}$ for the integral in Eq.~\eqref{faddeev:eq}.
         In the following subsection we collect the ingredients of Eqs.~(\ref{faddeev:eq}--\ref{KSS}), i.e. the dressed-quark propagator $S$
         and the quark-quark kernel $\mathcal{K}$.

         \begin{figure}[t]
                    \begin{center}

                    \includegraphics[scale=0.13]{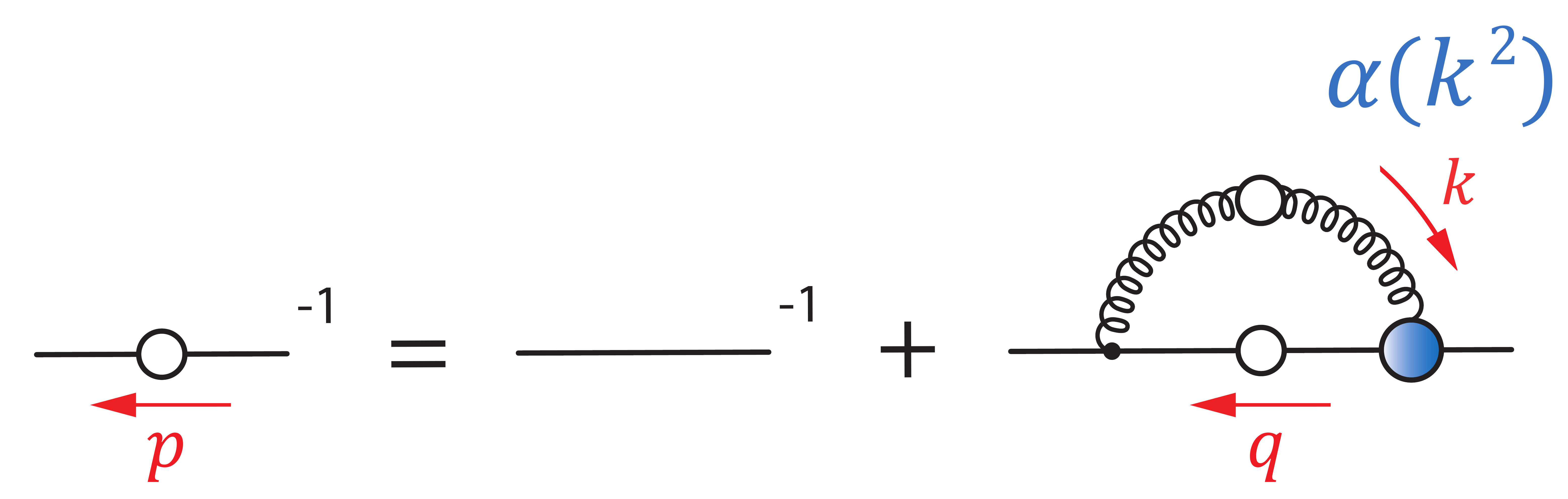}
                    \caption{(Color online) Quark DSE~\eqref{quarkdse} in rainbow-ladder truncation.
                             }\label{fig:dse}

                    \end{center}
        \end{figure}


\subsection{Quark propagator and $qq$ kernel} \label{sec:qprop}

          The fundamental ingredient which appears in the covariant Faddeev equation and provides a link between hadron properties
          and the underlying structure of QCD is the dressed quark propagator $S(p)$.
          It is expressed in terms of two scalar functions, the quark wave-function renormalization $1/A(p^2)$ and the quark mass function $M(p^2)$:
                  \begin{equation}\label{qprop}
                      S^{-1}(p) = A(p^2)\,\left( i\Slash{p} + M(p^2) \right).
                  \end{equation}
          Dynamical chiral symmetry breaking is manifest in a non-perturbative enhancement
          of both dressing functions $M(p^2)$ and $A(p^2)$ at small momenta which indicates
          the dynamical generation of a large constituent-quark mass.
          Such an enhancement emerges in the solution of the quark DSE, cf. Fig.~\ref{fig:dse}:
                  \begin{equation}\label{quarkdse}
                      S^{-1}_{\alpha\beta}(p) = Z_2 \left( i\Slash{p} + m \right)_{\alpha\beta}  +
      		               \int\limits_q \mathcal{K}_{\alpha\alpha'\beta'\beta}(p,q) \,S_{\alpha'\beta'}(q)\,,
                  \end{equation}
          where $Z_{2}$ is the quark renormalization constant. The bare current-quark mass $m$
          constitutes an input of the equation and can be readily varied from the chiral limit up to the heavy-quark regime.
      	  The interaction kernel $\mathcal{K}$ includes the dressed gluon propagator as well as one bare
      	  and one dressed quark-gluon vertex.

         \begin{figure}[tbp]
                    \begin{center}

                    \includegraphics[scale=0.40]{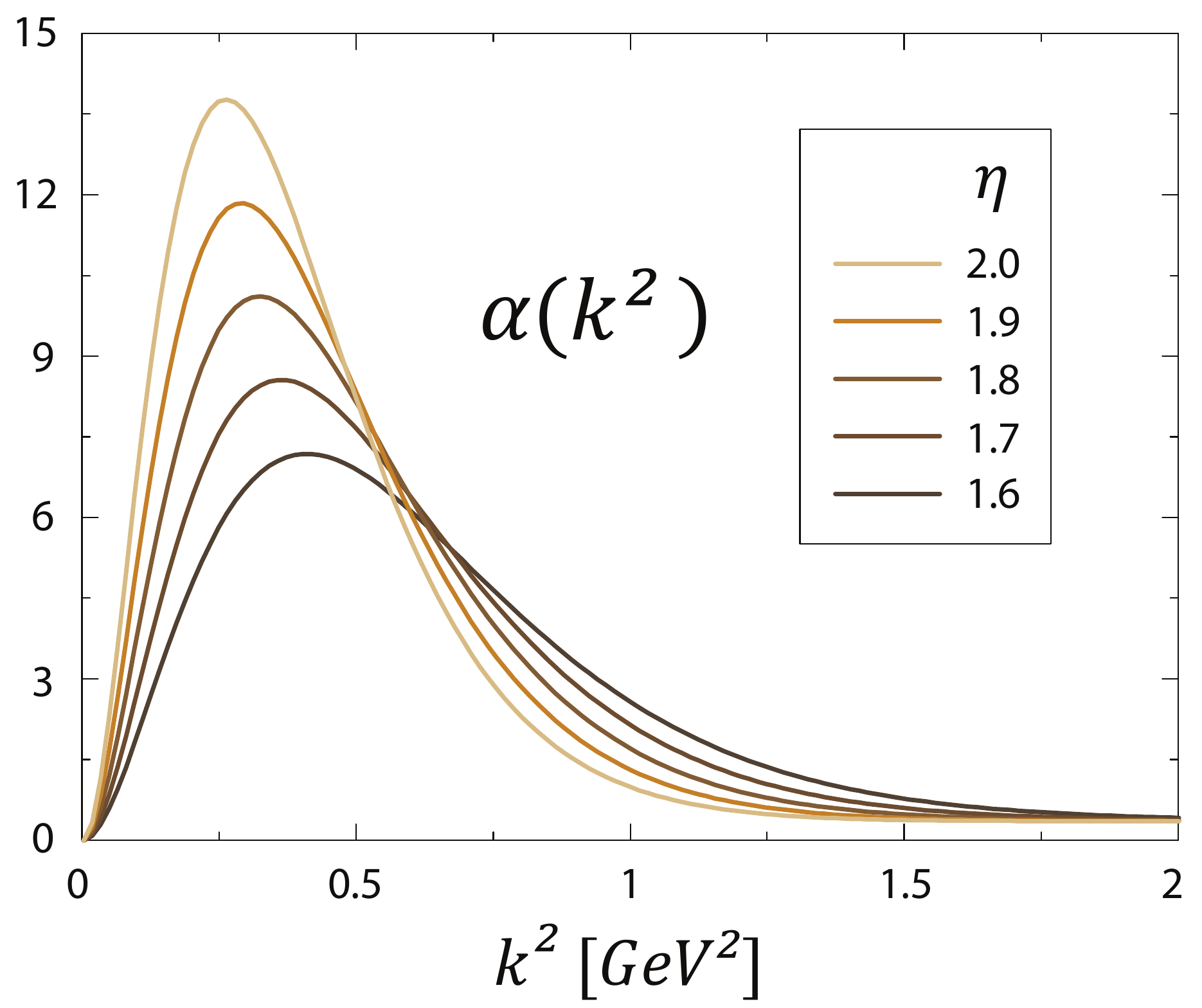}
                    \caption{(Color online) Effective coupling $\alpha(k^2)$ of Eq.\,\eqref{couplingMT}, evaluated for $\Lambda = 0.72$ GeV
                                            corresponding to the $u/d-$quark mass and in the range $\eta \in [1.6,2.0]$.}\label{fig:coupling}

                    \end{center}
        \end{figure}

        While in principle the dressed gluon propagator and quark-gluon vertex could
    	be obtained as solutions of the infinite coupled tower of QCD's DSEs,
    	practical numerical studies rely upon truncations, combined with
        substantiated ans\"atze for Green functions which are not solved for explicitly.
        The pion's nature as the Goldstone boson of spontaneous chiral symmetry breaking
        necessitates a truncation that preserves the axial-vector Ward-Takahashi identity.
        The latter relates the kernel of the quark DSE with that of a meson BSE,
        ensures a massless pion in the chiral limit and leads to a generalized Gell-Mann--Oakes--Renner relation~\cite{Maris:1997hd,Holl:2004fr}.
            Such a symmetry-preserving truncation scheme was described in~\cite{Munczek:1994zz,Bender:1996bb},
            and its lowest order is
            the rainbow-ladder (RL) truncation which amounts to an iterated dressed-gluon exchange between quark and antiquark.
        In rainbow-ladder only the vector part $\sim \gamma^\mu$ of the quark-gluon vertex is retained whose
        non-perturbative dressing, together with that of the gluon propagator, is absorbed into an effective coupling $\alpha(k^2)$ which is modeled.
    	The kernel $\mathcal{K}$ of both quark DSE and Faddeev equation then reads:
                \begin{equation}\label{RLkernel}
                    \mathcal{K}_{\alpha\alpha'\beta\beta'} =  Z_2^2 \, \frac{ 4\pi \alpha(k^2)}{k^2} \, T^{\mu\nu}_k \gamma^\mu_{\alpha\alpha'} \,\gamma^\nu_{\beta\beta'},
                \end{equation}
        where $T^{\mu\nu}_k=\delta^{\mu\nu} - \hat{k}^\mu \hat{k}^\nu$
        is a transverse projector with respect to the gluon momentum $k$ and a hat
    	denotes a normalized 4-vector.

        At large gluon momenta the effective coupling $\alpha(k^2)$ is constrained by perturbative QCD, whereas
        the details of the interaction in the deep infrared are not important for hadronic ground states~\cite{Blank:2010pa}.
        At small and intermediate momenta it must exhibit sufficient strength to
        allow for dynamical chiral symmetry breaking and the dynamical
    	generation of a constituent-quark mass scale.
        We employ the frequently used ansatz~\cite{Maris:1999nt}
                        \begin{equation}\label{couplingMT}
                            \alpha(k^2) = \pi \eta^7  \left(\frac{k^2}{\Lambda^2}\right)^2 \!\! e^{-\eta^2 \left(\frac{k^2}{\Lambda^2}\right)} + \alpha_\text{UV}(k^2) \,,
                        \end{equation}
            where the second term reproduces the logarithmic decrease of QCD's perturbative running coupling and vanishes at $k^2=0$.
            The first term supplies the necessary infrared strength and is characterized by two parameters: 
            an infrared scale $\Lambda$ and a dimensionless width parameter $\eta$, cf. Fig.~\ref{fig:coupling}.

        In combination with the interaction of Eq.~\eqref{couplingMT}, the RL truncation
        has been extensively used in Dyson-Schwinger studies of hadrons.
        Upon setting the scale $\Lambda = 0.72$~GeV to reproduce the experimental pion decay constant,
        RL provides a reasonable description of pseudoscalar-meson, vector-meson,
        nucleon and $\Delta$ ground-state properties, see e.g.~\cite{Maris:2005tt,Maris:2006ea,Krassnigg:2009zh,Nicmorus:2008vb} and references therein.
        Moreover, these observables have turned out to be largely insensitive to the shape of the coupling in the infrared~\cite{Maris:1999nt,Krassnigg:2009zh}; i.e.,
        to a variation of the parameter $\eta$ around the value $\eta \approx 1.8$.

        Progress has also been made for other meson quantum numbers such as axial-vector and pseudoscalar isosinglet mesons
        whose properties are subject to substantial corrections beyond rainbow-ladder~\cite{Watson:2004kd,Alkofer:2008tt,Alkofer:2008et,Fischer:2008wy,Fischer:2009jm,Chang:2009zb,Chang:2010jq}.
        However, such analyses typically require a significant amplification of numerical effort
        which is not yet feasible for studies in the baryon sector.
        Important attractive contributions beyond RL come from a pseudoscalar meson-cloud which augments
        the 'quark core' of dynamically generated hadron observables in the chiral regime and vanishes with increasing current-quark mass.
        Such effects are missing in a RL truncation.
        Thus, the present work aims at investigating the electromagnetic form factors of the nucleon's quark core. 

        We note that through Eq.~\eqref{couplingMT} all parameters of the interaction $\alpha(k^2)$, and thereby
        all equations that appear in subsequent sections, are fixed by using information from pion properties only.


    \renewcommand{\arraystretch}{1.0}

\subsection{Nucleon amplitude}

            Upon having determined the input of the covariant Faddeev equation through Eqs.~(\ref{qprop}--\ref{couplingMT})
            one can proceed with its solution.
            First results for the nucleon's mass and bound-state amplitude were reported in Ref.~\cite{Eichmann:2009qa,Eichmann:2009en}.
            While the mass can be reliably determined with relatively modest numerical accuracy,
            the form-factor computation requires a significantly higher resolution of the Faddeev amplitude, especially at larger photon momentum transfer.
            In the present work we use a solution technique that exploits the permutation-group properties of the amplitude.
            This enables us to drastically reduce the involved CPU times and solve the Faddeev equation with its full momentum dependence.
            The method is described in App.~\ref{sec:faddeeveqsolution}.

            The structure of the nucleon amplitude, together with its basis decomposition and permutation-group properties, is discussed in detail in App.~\ref{sec:nucamp}.
            In the following we will highlight some key aspects.
            The spin-flavor structure of the on-shell nucleon amplitude can be expressed as
            \begin{equation}\label{amplitude:spin-flavor}
                \mathbf{\Psi} = \Psi\cdot\mathsf{F} = \sum_{n=1}^2 \Psi_n\,\mathsf{F}_n\,,
            \end{equation}
            where the Dirac amplitude $\Psi$ and the isospin-$\nicefrac{1}{2}$ flavor tensor $\mathsf{F}$ of Eq.~\eqref{FAD:flavor}
            transform as doublets under the permutation group $\mathds{S}^3$,
            with mixed-antisymmetric entries $\Psi_1$, $\mathsf{F}_1$ and
            mixed-symmetric components $\Psi_2$, $\mathsf{F}_2$, respectively.
            The structure of Eq.~\eqref{amplitude:spin-flavor} ensures the Pauli principle: the nucleon amplitude involving its full spin-flavor-color structure
            must be antisymmetric under quark exchange, hence its spin-flavor part has to be symmetric.
            The spinor parts $\Psi_n$ involve 64 covariant, orthogonal and momentum-dependent Dirac structures $\mathsf{X}_{k,ij\omega}$,
            \begin{equation}\label{amplitude:reconstruction1}
                \Psi_n(p,q,P) = \sum_{kij\omega} f_{n,kij\omega}\,\mathsf{X}_{k,ij\omega}\,, 
            \end{equation}
            which are discussed in detail in App.~\ref{sec:orthonormalbasis}.
            The amplitude dressing functions $f_{n,kij\omega}$ depend on the 5 Lorentz-invariant
            momentum variables 
            \begin{equation} \label{mom-variables}
                p^2\,, \quad q^2\,,\quad   z_0=\widehat{p_T}\cdot\widehat{q_T} \,,\quad z_1 = \hat{p}\cdot\hat{P} \,,\quad  z_2 = \hat{q}\cdot\hat{P}\,,
            \end{equation}
            where a hat denotes a normalized 4-vector and the subscript '$T$' a transverse projection with respect to the
            nucleon momentum $P$.
            The total momentum-squared is fixed: $P^2=-M^2$.

            The orthogonal basis elements $\mathsf{X}_{k,ij\omega}$ are eigenstates of
            total quark spin and orbital angular momentum in the nucleon's rest frame;
            the corresponding partial-wave decomposition is explained in App.~\ref{sec:partialwave}.
            The rest-frame nucleon amplitude is dominated by $s-$wave components, i.e. by the subset of eight relative-momentum independent
            basis elements which carry total quark spin $s=\nicefrac{1}{2}$ and orbital angular momentum $l=0$. We denote them here by
            \begin{equation}\label{s-wave-basis-elements}
            \begin{split}
                \mathsf{S}_\pm &:= \mathsf{X}_{1,11\pm} =  \Lambda_\pm \gamma_5 C \otimes \Lambda_+\,, \\
                \mathsf{V}_\pm &:= \mathsf{X}_{1,21\pm} =  \textstyle{\frac{1}{\sqrt{3}}}\,\gamma^\alpha_T\,\Lambda_\pm \gamma_5 C \otimes \gamma^\alpha_T\,\Lambda_+\,, \\[2mm]
                \mathsf{P}_\pm &:= \mathsf{X}_{2,11\pm} = \left( \gamma_5 \otimes \gamma_5 \right) \mathsf{S}_\pm\,, \\
                \mathsf{A}_\pm &:= \mathsf{X}_{2,21\pm} = \left( \gamma_5 \otimes \gamma_5 \right) \mathsf{V}_\pm\,,
            \end{split}
            \end{equation}
            where the $\gamma-$matrices $\gamma_T^\alpha$ are transverse with respect to the nucleon momentum $P$.
            The remaining basis elements are either $p-$ or $d-$waves.
            Table~\ref{tab:spd} shows the $s-$, $p-$ and $d-$ wave contributions to the nucleon's canonical normalization integral~\eqref{canonical-norm} at different current-quark masses.
            The $s-$wave elements contribute roughly $\nicefrac{2}{3}$ to the norm and the $p-$waves the remaining third.
            The $p-$wave contribution decreases, albeit very slowly, with higher quark mass
            which signals a substantial amount of orbital angular momentum in the nucleon's rest-frame amplitude
            well beyond the strange-quark mass.  

         \begin{figure}[t]
                    \begin{center}

                    \includegraphics[scale=0.35]{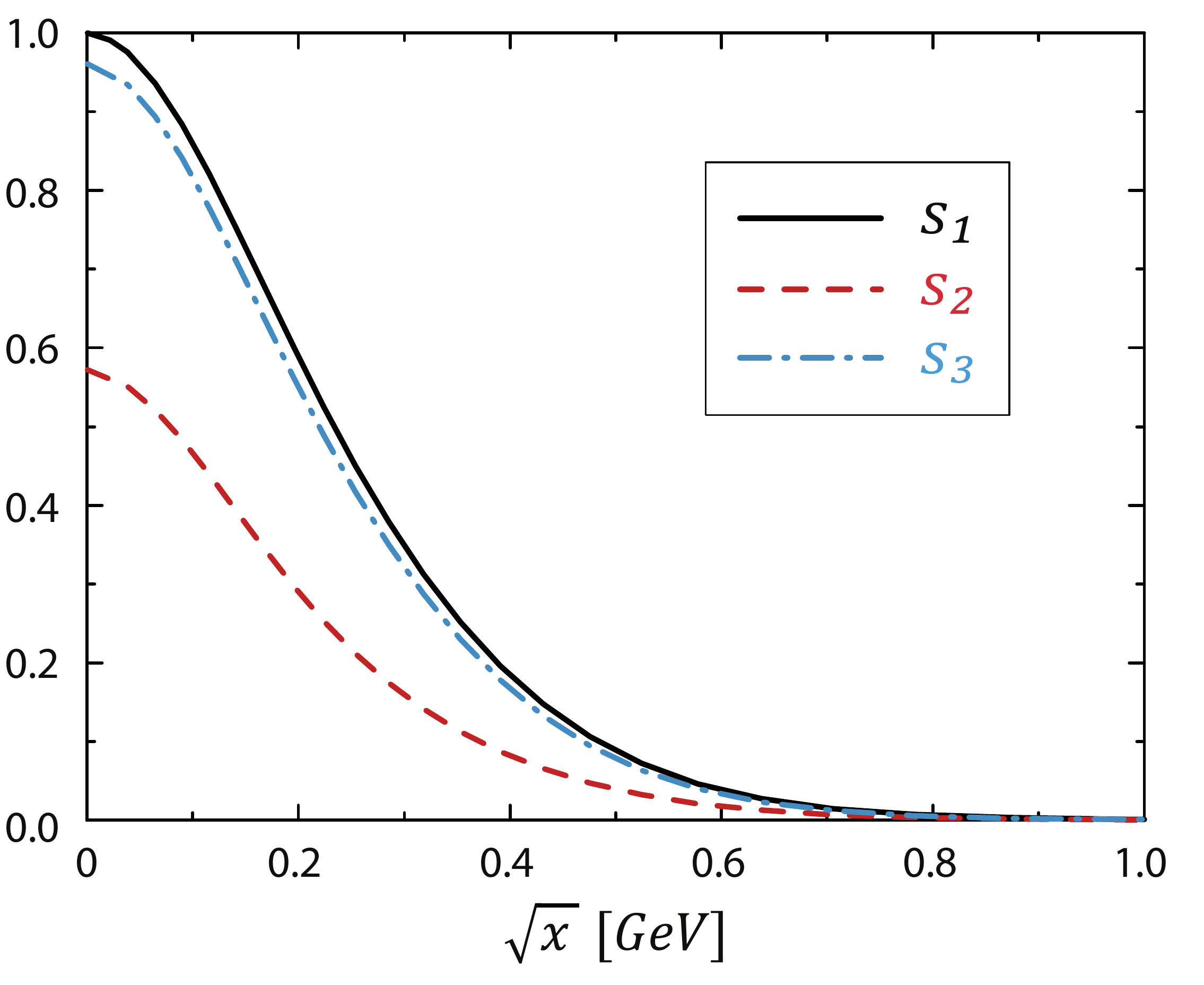}
                    \caption{(Color online) Result for the three dominant $s-$wave contributions in the nucleon's Faddeev amplitude.
                                            The plot shows the zeroth Legendre and Chebyshev moments (in the variables $y_1$, $z_0$ and $z_1$, $z_2$, respectively)
                                            of the dressing functions $s_i$ defined via Eqs.~\eqref{basis-doublets} and~\eqref{amplitude-symmetric}.}\label{fig:amplitude}

                    \end{center}
        \end{figure}

            To analyze the $s-$wave components in the nucleon amplitude in more detail it
            is instructive to rearrange the eight basis elements of Eq.~\eqref{s-wave-basis-elements} in permutation-group multiplets.
            This yields the orthonormal doublets
             \begin{equation}\label{basis-doublets}
             \begin{split}
                 \Psi^{(1)} &= \left(\begin{array}{c} \mathsf{S}_+ \\ \mathsf{A}_+\end{array}\right), \quad
                 \Psi^{(2)} = \frac{1}{\sqrt{3}}\left(\begin{array}{c} 2\,\mathsf{P}+\mathsf{S_-} \\ 2\,\widetilde{\mathsf{V}}-\mathsf{A}_-\end{array}\right), \\[2mm]
                 & \quad \Psi^{(3)} = \frac{1}{\sqrt{3}}\left(\begin{array}{c} \mathsf{S}_--\mathsf{P}+\sqrt{3}\,\mathsf{V} \\ \mathsf{A}_-+\widetilde{\mathsf{V}}+\sqrt{3}\,\widetilde{\mathsf{P}}\end{array}\right),
             \end{split}
             \end{equation}
            where the upper entries are mixed-antisymmetric with respect to the first two Dirac indices
            and the lower entries mixed-symmetric, and two further singlets
             \begin{equation}\label{basis-singlets}
             \begin{split}
                 \Psi_\mathcal{A} &=\frac{1}{\sqrt{3}} \left( \mathsf{S}_--\mathsf{P}-\sqrt{3}\,\mathsf{V} \right), \\
                 \Psi_\mathcal{S} &=\frac{1}{\sqrt{3}} \left( \mathsf{A}_-+\widetilde{\mathsf{V}}-\sqrt{3}\,\widetilde{\mathsf{P}} \right)
             \end{split}
             \end{equation}
             which are fully antisymmetric or symmetric, respectively.
             Here we defined $\mathsf{P} := (\mathsf{P}_+ + \mathsf{P}_-)/2$, $\widetilde{\mathsf{P}} := (\mathsf{P}_+ - \mathsf{P}_-)/2$
             and accordingly for $\mathsf{V}$ and $\widetilde{\mathsf{V}}$.

             Eqs.~(\ref{basis-doublets}--\ref{basis-singlets}) imply that, without including a dependence on the relative momenta,
             only three fully symmetric 
             Dirac-flavor combinations $\Psi^{(i)}\cdot\mathsf{F}$ can arise in the $s=\nicefrac{1}{2}$, $l=0$  subspace.
             They appear in combination with symmetric singlet dressing functions which
             are linear combinations of those associated with the basis elements in Eq.~\eqref{s-wave-basis-elements}
             and must depend on symmetric combinations of the momentum variables in Eq.~\eqref{mom-variables}.
             Denoting them by $s_i$, a fully symmetric spin/momentum-flavor amplitude is then obtained via
             \begin{equation}\label{amplitude-symmetric}
                 \mathbf{\Psi} = \sum_{i=1}^3 s_i\,  \Psi^{(i)}\cdot\mathsf{F} + \dots\,,
             \end{equation}
             where the dots refer to further combinations of Eqs.~(\ref{basis-doublets}--\ref{basis-singlets}) with mixed-(anti-)symmetric dressing functions,
             and also to the remaining $p-$ and $d-$wave components.
             In Eq.~\eqref{doublet-variables} we define momentum variables that transform as multiplets under $\mathds{S}^3$,
             namely a symmetric singlet variable
             \begin{equation}
                 x := \frac{p^2}{4} + \frac{q^2}{3}\,,
             \end{equation}
             and four dimensionless angular variables $y_1$, $y_2$, $w_1$, $w_2$ which form doublets.
             The dressing functions $s_i$ can then only depend on the variable $x$
             and the symmetric combinations
             $y_1^2+y_2^2$, $w_1^2+w_2^2$, and $y_1 w_1+y_2 w_2$.

             \begin{table}[t]

                \begin{center}
                \begin{tabular}{   c @{\;\;} ||  @{\;\;}c@{\;\;} | @{\;\;}c@{\;\;} | @{\;\;}c@{\;\;}    }

                    $m_\pi$\,[GeV]          &  $0.14$         &     $0.34$   &  $0.75$       \\   \hline

                    $s-$wave         &  $0.66$        &     $0.67$    &   $0.69$      \\
                    $p-$wave         &  $0.33$        &     $0.32$    &   $0.30$      \\
                    $d-$wave         &  $0.01$        &     $0.01$    &    $0.01$

                \end{tabular} \caption{$s-$, $p-$ and $d-$wave contributions to the nucleon's canonical normalization at three pion masses,
                                       expressed as fractions of $1$. The first column corresponds to the physical $u/d-$quark mass.
                                       }\label{tab:spd}
                \end{center}

        \end{table}

             The full solution of the Faddeev equation indeed reveals the three singlet dressing functions $s_i$
             to contribute the bulk to the $s-$wave fraction in the normalization.
             Their angular dependence is weak, especially in the variables $z_2$ and $z_0$, and a corresponding
             polynomial expansion vanishes rapidly.
             The zeroth angular moments 
             of the three $s_i$ are plotted in Fig.~\eqref{fig:amplitude} as a function of the variable $\sqrt{x}$.
             All three dressing functions turn out to be large; in particular, $s_1$ and $s_3$ are almost identical in size.

\begin{figure}[t]
                    \begin{center}

                    \includegraphics[scale=1.1]{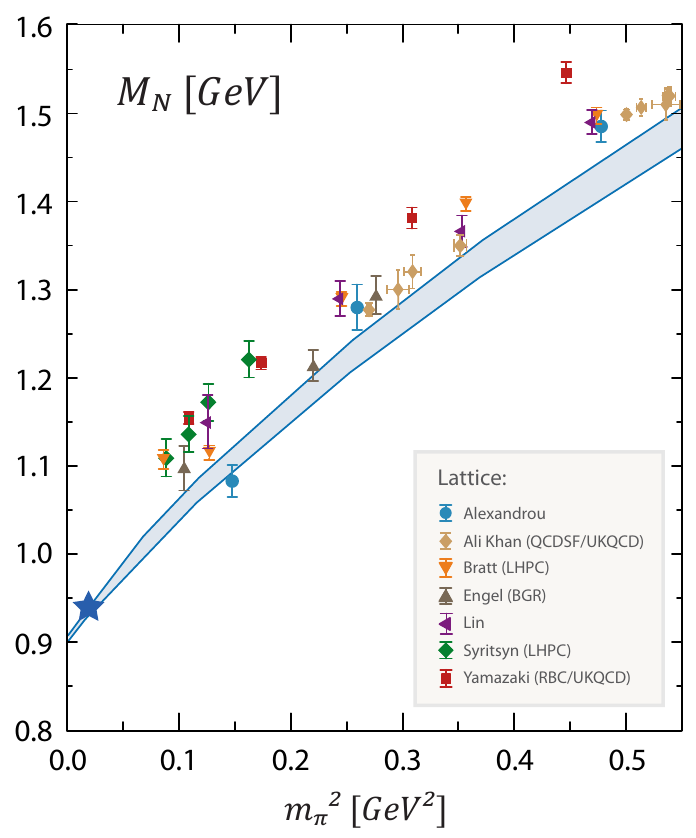}
                    \caption{(Color online) Current-mass evolution of the nucleon mass obtained from the covariant Faddeev equation.
                             The band corresponds to a variation of the width parameter $\eta=1.8\pm 0.2$;
                             the star denotes the experimental value.
                             We compare to a selection of lattice data from Refs.~\cite{Ali_Khan:2003cu,Alexandrou:2006ru,Syritsyn:2009mx,Yamazaki:2009zq,Bratt:2010jn,Lin:2010fv,Engel:2010my}.}\label{fig:nucleonmass}

                    \end{center}
        \end{figure}

        The resulting current-mass evolution of the nucleon's mass is displayed in Fig.~\ref{fig:nucleonmass}.
        The pion mass was obtained from its pseudoscalar-meson Bethe-Salpeter equation with the same rainbow-ladder input.
        The scale $\Lambda$ in Eq.~\eqref{couplingMT} was fixed to reproduce the experimental pion decay constant.
        In agreement with previous meson and quark-diquark studies, the sensitivity to the infrared shape of the effective coupling $\alpha(k^2)$ is small;
        this is indicated by the band which corresponds to a variation $\eta=1.8\pm 0.2$, cf. Fig.~\ref{fig:coupling}.
        At the physical $u/d-$quark mass, our result $M_N=0.94$ GeV 
        is in excellent agreement with the experimental value, and its current-mass evolution compares reasonably well with lattice data at higher quark masses.

        In connection with Fig.~\ref{fig:nucleonmass} we reiterate that contributions from a pseudoscalar-meson cloud are absent in a rainbow-ladder truncation;
        the current approach can therefore be viewed to describe a hadronic quark core.
        Such corrections can be estimated from chiral effective field theory and would yield a reduction of $\sim 20-30\%$ of
        the nucleon's core mass in the chiral region~\cite{Nicmorus:2008vb}. The proximity between our calculated mass and the experimental and lattice values
        therefore suggests a non-perturbative cancelation mechanism beyond rainbow-ladder.
        Indeed, such a behavior emerges for the $\rho-$meson
        where attractive pion-cloud effects beyond RL are essentially saturated
        by further repulsive contributions from the quark-gluon vertex and the quark-antiquark kernel~\cite{Fischer:2008wy,Fischer:2009jm,Chang:2009zb,Chang:2010jq}.
        In addition, the second type of corrections dominates in scalar and axial-vector mesons 
        which explains why these quantum numbers are not well reproduced in a RL truncation.
        Given the qualitatively quite similar behavior of the $\rho-$meson mass in the present framework
        in comparison with lattice data, as well as nucleon and $\Delta$ masses in a quark-diquark calculation~\cite{Eichmann:2008ef,Nicmorus:2010mc},
        it is conceivable that a similar mechanism might be at work here.

        We note that the Faddeev result for the nucleon mass in Fig.~\eqref{fig:nucleonmass} is in excellent agreement with that obtained from the
        quark-diquark calculation (cf. Fig.~2 in Ref.~\cite{Nicmorus:2010mc}), where the $qq$ scattering matrix was reduced to a sum of scalar and axial-vector diquark channels
        which were solved from their Bethe-Salpeter equations.
        This highlights the importance of the diquark concept as relevant degrees of freedom within hadrons,
        and it specifically shows that including only the lightest diquarks can already account for the bulk of the nucleon mass.
        In Section~\ref{sec:results} it will turn out that this observation also holds for the overall behavior of the nucleon's electromagnetic form factors.

         \begin{figure*}[htp]
                    \begin{center}
                    \includegraphics[scale=0.105]{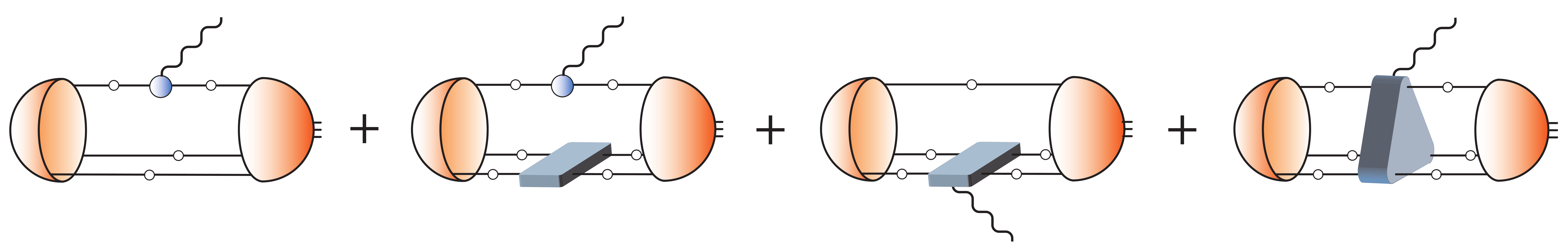}
                    \caption{(Color online) General expression for a baryon's electromagnetic current (modulo permutations)
                             in the three-quark framework.  }\label{fig:faddeev-current}
                    \end{center}
        \end{figure*}


\section{Electromagnetic current} \label{sec:ffs}

 \renewcommand{\arraystretch}{1.2}

\subsection{Construction of the current}

             Having numerically computed the nucleon amplitude, we proceed with
             the construction of the nucleon's electromagnetic current.
             Its general structure is given by (cf. App.~\ref{sec:currentgeneral})
            \begin{equation}\label{eq:current2}
                J^\mu(P,Q) = i \Lambda_+^f \left( F_1  \gamma^\mu - \frac{F_2}{2M}\, \sigma^{\mu\nu} Q^\nu \right) \Lambda_+^i\,
            \end{equation}
            and parameterized by two form factors: the Dirac form factor $F_1(Q^2)$ and the Pauli form factor $F_2(Q^2)$.
            If $P_i$ and $P_f$ denote the initial and final nucleon momenta,
             the exchanged photon momentum is given by $Q=P_f-P_i$ and the average total momentum by $P=(P_i+P_f)/2$.
             The positive-energy projectors $\Lambda_+^{f,i} = \Lambda_+(P_{f,i})$ are defined in Eq.~\eqref{energy-projectors}.
             The Dirac and Pauli form factors can be expressed through the Sachs form factors:
              \begin{equation} \label{ff:sachs}
                      G_E = F_1 - \tau F_2\,, \quad
                      G_M = F_1 + F_2\,,
              \end{equation}
              with $\tau=Q^2/(4M^2)$.
             Their static dimensionless values are the proton's and neutron's charges $G_{E}^{p,n}(0) =\{  1,0 \}$
             and their magnetic dipole moments $G_{M}^{p,n}(0) = \mu^{p,n}$.

            In order to relate the electromagnetic current of Eq.~\eqref{eq:current2}
            to the underlying description of the nucleon as a composite object,
            one must specify how a photon couples to its constituents.
            A systematic construction principle for the nucleon-photon current
            based on electromagnetic gauge invariance is the 'gauging of equations' prescription~\cite{Haberzettl:1997jg,Kvinikhidze:1998xn,Kvinikhidze:1999xp} which was previously
            used to derive the electromagnetic current in the quark-diquark model~\cite{Oettel:1999gc,Oettel:2000ig}.
            We will briefly sketch the procedure here.
             Gauging, indicated by an index $\mu$, has the properties of a derivative, i.e. it is linear and satisfies Leibniz' rule.
             In the present context, $\mu$ denotes the coupling to a photon with momentum $Q$:
             when acting upon an $n-$point Green function, it yields an $(n+1)-$point function with an additional photon leg.
             For instance, the quark-photon vertex is the gauged inverse quark propagator: $\Gamma^\mu_\text{q} = (S^{-1})^\mu$,
             and hence $S^\mu = -S \,(S^{-1})^\mu\, S = -S \,\Gamma^\mu_\text{q} \,S$.

             To derive a baryon's electromagnetic current in this framework,
             we recall the nonperturbatively resummed Dyson series for the three-body scattering matrix $T$,
             i.e. Dyson's equation, which is the starting point of the bound-state approach:
             \begin{equation}\label{dyson-eq}
                  T = K + K \,G_0 \,T \quad \Longleftrightarrow \quad T^{-1} = K^{-1} - G_0\,,
             \end{equation}
             where $G_0$ is the product of three dressed quark propagators and $K = \widetilde{K}\,G_0^{-1}$ is the three-body kernel of Eq.~\eqref{three-quark-eq}.
             A pole in the scattering matrix defines a bound state on its mass shell $P^2=-M^2$.
             The scattering matrix at the pole assumes the form
             \begin{equation}\label{poles-in-T}
                T\stackrel{P^2=-M^2}{\longlonglongrightarrow} \frac{\Psi\,\conjg{\Psi}}{P^2+M^2}\,,
             \end{equation}
             where $\Psi$ is the baryon's bound-state amplitude and $\conjg{\Psi}$ its charge conjugate,
             and we suppressed all indices as well as the momentum dependence and dimensionful prefactors.
             At the pole, Dyson's equation reduces to the homogeneous bound-state equation~\eqref{three-quark-eq} for the amplitude $\Psi$:
             \begin{equation}\label{boundstate-eq}
                 K \,G_0\,\Psi = \Psi \quad \Longleftrightarrow \quad T^{-1} \Psi = 0\,.
             \end{equation}
             Evaluating the relation $T' = -T \,(T^{-1})' \,T$ at the bound-state pole,
             where $'$ denotes the derivative $d/dP^2$, yields in combination with  Eq.~\eqref{poles-in-T} the
             on-shell canonical normalization condition:
             \begin{equation}\label{canonical-norm}
                \conjg{\Psi} \left(T^{-1}\right)' \Psi = 1 \,.
             \end{equation}

             The electromagnetic current matrix is the residue of the gauged scattering matrix $T^\mu$ at the bound-state pole:
             \begin{equation}
                 T^\mu \stackrel{P_i^2=P_f^2=-M^2}{\longlonglonglongrightarrow} - \frac{\Psi_f\,J^\mu \,\conjg{\Psi}_i}{(P_f^2+M^2)(P_i^2+M^2)}\,,
             \end{equation}
             where $\Psi_i = \Psi(p_i,q_i,P_i)$ and $\Psi_f = \Psi(p_f,q_f,P_f)$ are in- and outgoing amplitudes with different
             momentum dependencies.
             On the other hand, Eq.~\eqref{poles-in-T} yields
             \begin{equation}
             \begin{split}
                  T^\mu   = & -T\left(T^{-1}\right)^\mu T \stackrel{P_i^2=P_f^2=-M^2}{\longlonglonglongrightarrow} \\
                       &    -\frac{\Psi_f\,\conjg{\Psi}_f}{P_f^2+M^2}\left(T^{-1}\right)^\mu\frac{\Psi_i\,\conjg{\Psi}_i}{P_i^2+M^2}\,,
             \end{split}
             \end{equation}
             and by comparing these two equations one obtains the electromagnetic current as
             the gauged inverse scattering matrix element between the onshell bound-state amplitudes:
             \begin{equation}\label{emcurrent-gauging}
                J^\mu = \conjg{\Psi}_f \left(T^{-1}\right)^\mu \Psi_i \,.
             \end{equation}
             Electromagnetic current conservation,
             \begin{equation}
                 Q^\mu J^\mu = \conjg{\Psi}_f \,Q^\mu \left(T^{-1}\right)^\mu \Psi_i=0\,,
             \end{equation}
             follows from a Ward-Takahashi identity for the five-point function $(T^{-1})^\mu$ upon exploiting the onshell condition
             $T^{-1}\,\Psi=0$ from Eq.~\eqref{boundstate-eq}. In the limit $Q^2\rightarrow 0$, Eq.~\eqref{emcurrent-gauging} reproduces the canonical normalization condition,
             and charge conservation $G_E^p(0)=1$ is automatically satisfied if the Faddeev amplitude
             is canonically normalized.

             To extract the specific form of the current in the three-body approach,
             one derives from Eq.~\eqref{dyson-eq}: 
             \begin{equation}
                 \left(T^{-1}\right)^\mu = \left(K^{-1}\right)^\mu - G_0^\mu = -K^{-1}\,K^\mu\,K^{-1} - G_0^\mu\,,
             \end{equation}
             which, upon sandwiching between onshell nucleon amplitudes and implementing the bound-state relation~\eqref{boundstate-eq},
             yields the following expression for the electromagnetic current:
             \begin{equation} \label{emcurrent-gauging-2}
                J^\mu = - \conjg{\Psi}_f \Big[ G_0^\mu + G_0\,K^\mu\,G_0 \Big] \Psi_i\,.   
             \end{equation}
            The kernel $K = \widetilde{K}\,G_0^{-1}$ that appears in the bound-state equation~\eqref{three-quark-eq}
            of a baryon described by three valence quarks is given by
            \begin{equation}
                K = K_\text{irr} + \sum_{a=1}^3 K_{(a)} \, S^{-1}_{(a)},
            \end{equation}
            hence the corresponding gauged kernel reads
            \begin{equation}
                K^\mu = K_\text{irr}^\mu + \sum_{a=1}^3 \left( K_{(a)}^\mu \, S_{(a)}^{-1} +  K_{(a)} \,\Gamma^\mu_{\text{q},(a)} \right),
            \end{equation}
            where $\Gamma^\mu_{\text{q},(a)}$ is the dressed quark-photon vertex acting on quark $(a)$.
            In conjunction with Eq.~\eqref{emcurrent-gauging-2} this describes the general expression
            for a baryon's electromagnetic current which is displayed in Fig.~\ref{fig:faddeev-current}.
            It comprises an impulse-approximation diagram, a component involving the two-quark kernel,
            and two diagrams where the photon couples to the $qq$ and the $qqq$ kernels, respectively.
            In addition one must take into account the diagrams with permuted quark lines.

\begin{figure}[tb]
            \begin{center}
            \includegraphics[scale=0.11]{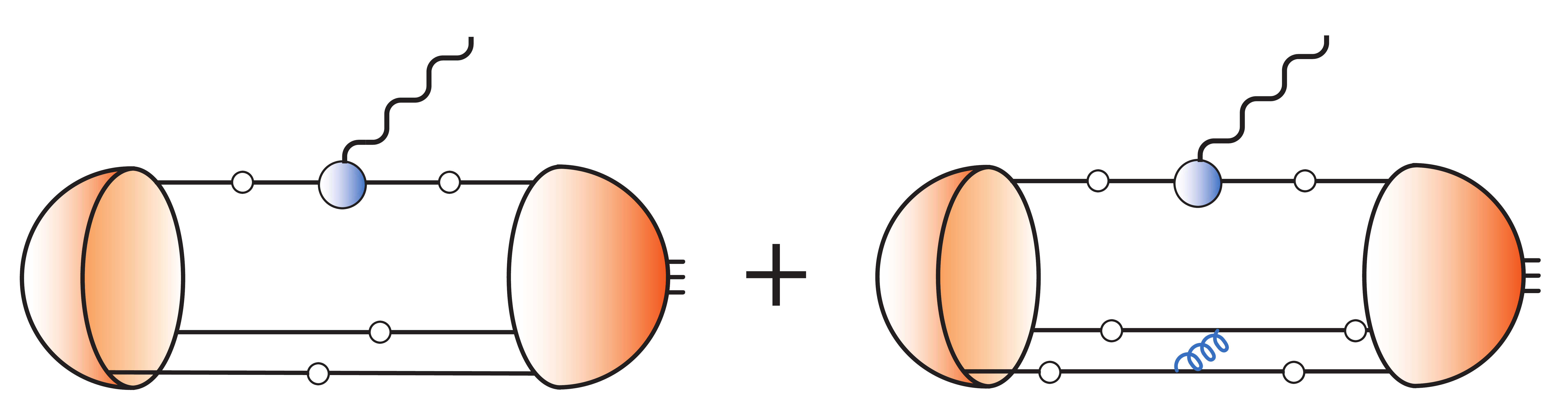}
            \caption{(Color online)
                     The two types of diagrams which contribute to the nucleon's three-body current in rainbow-ladder truncation, Eq.~\eqref{FADDEEV:Current}.} \label{fig:faddeev-current-specific}
            \end{center}
            \end{figure}

            The close relation between the three-body bound-state equation~\eqref{boundstate-eq}
            and the electromagnetic current~\eqref{emcurrent-gauging-2} makes clear that a given truncation must be respected by both.
            Neglecting the three-body irreducible contribution and employing a rainbow-ladder two-body kernel
            which is independent of the total momentum, hence $K_{(a)}^\mu=0$,
            finally yields the electromagnetic current
            \begin{equation}\label{FADDEEV:Current}
                J^\mu = -\conjg{\Psi}_{\!f} \left[ G_0^\mu + G_0 \left(\sum_{a=1}^3 K_{(a)} \, \Gamma^\mu_{\text{q},(a)}\right) G_0 \right] \Psi_i\,
            \end{equation}
            which is depicted in Fig.~\ref{fig:faddeev-current-specific} and employed in the present calculation.
            It is now reduced to the impulse-approximation diagram and a further piece that involves the rainbow-ladder kernel. 
            These contributions are worked out in detail in App.~\ref{sec:emcurrent-worked-out}.
            The only additional component which appears in Fig.~\ref{fig:faddeev-current-specific} is the dressed quark-photon vertex.
            It is computed consistently from its rainbow-ladder truncated inhomogeneous Bethe-Salpeter equation and described in App.~\ref{app:qpv}.

\subsection{Extraction of the form factors}

             By virtue of Eq.~\eqref{eq:current2},
             the electric and magnetic Sachs form factors are extracted from the electromagnetic current $J^\mu$ through the Dirac traces
             \begin{equation}\label{obtain-ffs}
             \begin{split}
                 G_E(Q^2) &= \frac{1}{2i\,\sqrt{1+\tau}}\, \text{Tr}\, \{ J^\mu \hat{P}^\mu \}, \\
                 G_M(Q^2) &= \frac{i}{4\tau}\,\text{Tr}  \,\{ J^\mu \gamma^\mu_T \},
             \end{split}
             \end{equation}
             where $\tau=Q^2/(4M^2)$, $P$ is the average momentum of incoming and outgoing nucleon, and $\gamma_T^\mu$ is transverse with respect to $P$.

             Only the diagrams illustrated in Fig.~\ref{fig:faddeev-current-specific},
             where the photon couples to the respective \textit{upper} quark lines, need to be computed explicitly;
             the remaining permuted diagrams can be obtained from the relation~\eqref{current:permutations}.
             The incoming and outgoing nucleon amplitudes are given by the respective combinations of mixed-antisymmetric and
             mixed-symmetric Dirac amplitudes $\Psi_1$ and $\Psi_2$ that appear in Eq.~\eqref{amplitude:spin-flavor}.
             Upon working out the flavor traces one finds that only diagrams contribute whose amplitudes
             in the incoming and outgoing state share the same symmetry. Thus, both electric and magnetic form factors can be decomposed as
             \begin{equation}
                 G = G^{AA} + G^{SS} = G^{AA}_\text{IMP} + G^{AA}_\text{K} + G^{SS}_\text{IMP} + G^{SS}_\text{K}\,,
             \end{equation}
             where $G^{AA}$ denotes a form factor obtained from two mixed-antisymmetric amplitudes and
             $G^{SS}$ the respective contribution with two mixed-symmetric amplitudes.

             The structure of the basis elements in Eqs.~(\ref{s-wave-basis-elements}--\ref{basis-doublets}) suggests
             to identify $G^{AA}$ as the predominantly scalar-diquark component and
             $G^{SS}$ as the axial-vector diquark contribution to the form factors.
             While this can serve intuition, 
             there is however no one-to-one mapping between $\Psi_1$, $\Psi_2$ and the scalar- and axial-vector
             diquark contributions to the nucleon amplitude in the quark-diquark model. Ultimately, $\Psi_1$ and $\Psi_2$
             carry contributions from all diquark channels.

             Summing all permuted diagrams, cf.~Eq.~\eqref{current-proton-neutron}, yields the proton and neutron form factors as the combinations
             \begin{equation}\label{ff-relations-from-aa-ss}
                 G^p = 2 \,G^{AA}\,, \quad G^n = G^{SS} - G^{AA}\,,
             \end{equation}
             where we again suppressed the subscripts $E$ and $M$ for the electric and magnetic contributions.
             Similarly, one obtains the isoscalar and isovector combinations
             \begin{equation}
             \begin{split}
                 G^s &= G^p + G^n = G^{AA} + G^{SS}\,, \\
                 G^v &= G^p - G^n = 3 \,G^{AA} - G^{SS}\,.
             \end{split}
             \end{equation}
             Finally, the flavor-separated up-/down-quark contributions in the proton which, owing to charge symmetry,
             equal the down-/up-quark contributions in the neutron, are given by
             \begin{equation}
             \begin{split}
                 G^u &= 2\,G^p +G^n = 3 \,G^{AA} + G^{SS}\,, \\
                 G^d &= G^p + 2 \,G^n = 2 \,G^{SS}\,.
             \end{split}
             \end{equation}
             It is apparent that charge conservation $G_E^p(0)=1$, $G_E^n(0)=0$ requires $G_E^{AA}(0) = G_E^{SS}(0) = \nicefrac{1}{2}$,
             and a neutron electric form factor $G_E^n(Q^2)\geq 0$ can only occur if $G_E^{AA} \leq G_E^{SS}$.
             Similarly, the neutron's Dirac and Pauli form factors obtained from Eq.~\eqref{ff:sachs} are negative if $F_{1,2}^{AA} > F_{1,2}^{SS}$.

             We further comment upon the relative importance of the two diagrams in Fig.~\ref{fig:faddeev-current-specific}.
             From the kinematics in the form-factor diagrams, Eqs.~(\ref{ff-momenta-1}--\ref{ff-momenta-2}),
             one infers that only the relative momenta $p_i$, $p_f$ of the incoming and outgoing amplitudes
             and the quark momenta $p_1$, $p_2$, $p_3^\pm$ carry a dependence on the
             photon momentum transfer $Q^2$ (in the case of $p_1$ and $p_2$, through the average momentum $P$).
             The relative momenta $q_i$, $q_f$ as well as the gluon momentum $k$ in the second diagram
             are $Q^2-$independent, hence both diagrams should follow a similar $Q^2-$evolution.
             Indeed we find that in the accessible photon-momentum range $Q^2 \lesssim 7$~GeV$^2$ the ratios
             \begin{equation}
                 R(Q^2) := -G_\text{K}(Q^2)/G_\text{IMP}(Q^2)
             \end{equation}
             depend only weakly on $Q^2$: for the magnetic form factors, $R_M^{p,n} \sim 0.3 \dots 0.5$
             and for the electric proton form factor, $R_E^p \sim 0.4 \dots 1$.
             For the electric neutron form factor charge conservation implies $G_\text{K}(0) = -G_\text{IMP}(0)$, hence $R_E^n(0)=1$.
             While the kernel contribution to $G_E^n$ stays positive at all $Q^2>0$,
             the impulse-approximation diagram is negative at $Q^2=0$ and becomes positive at $Q^2 \approx 1$ GeV$^2$
             above which it dominates the form factor.
             Both contributions in Fig.~\ref{fig:faddeev-current-specific} are therefore hard,
             and the diagram which involves the $qq$ kernel
             and necessarily augments the impulse-approximation diagram to ensure
             current conservation stays important at large momentum transfer.


       \begin{figure*}[tp]
                    \begin{center}

                    \includegraphics[scale=0.11]{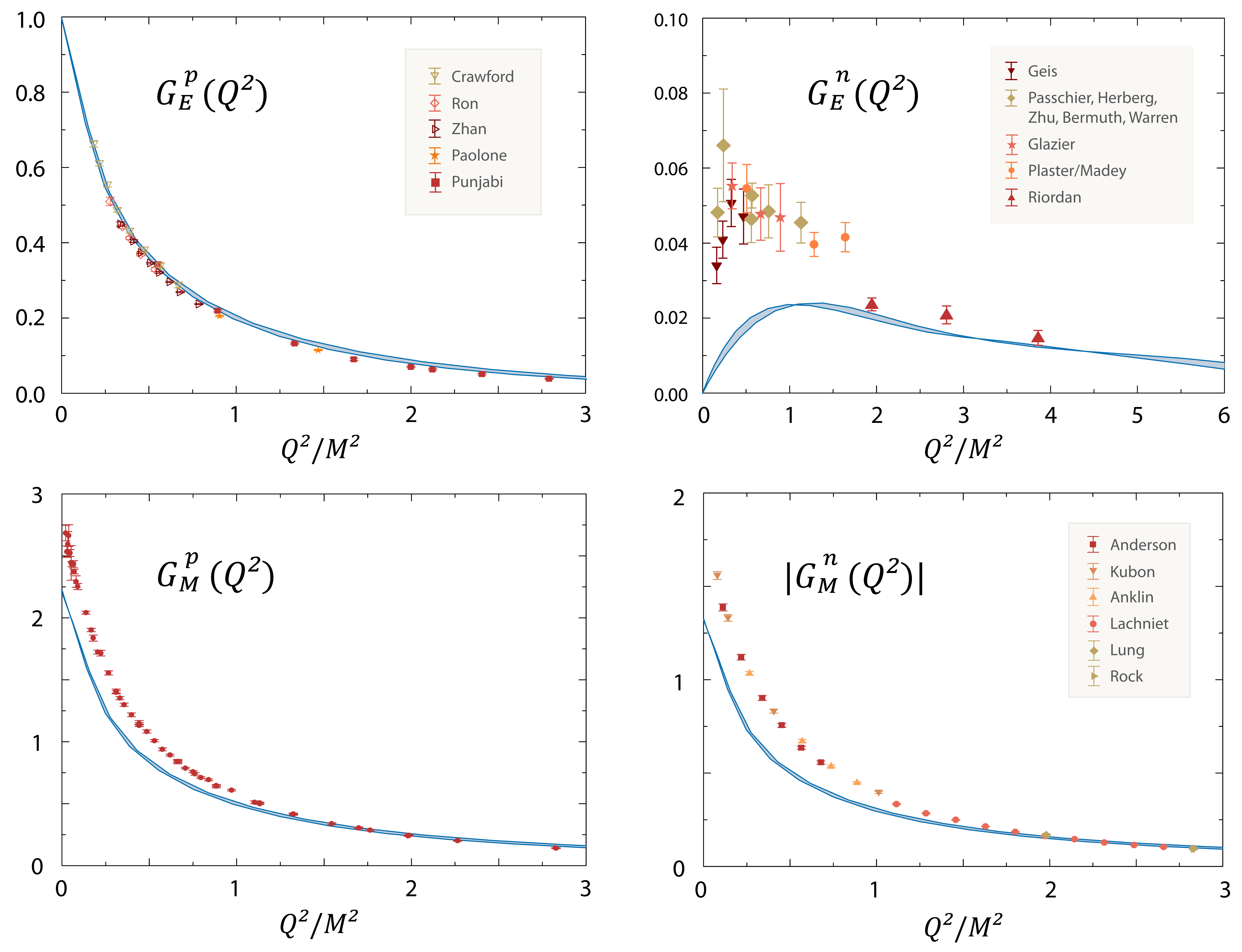}
                    \caption{(Color online) Electromagnetic form factors of the nucleon as a function of the photon momentum transfer $Q^2$.
                    The bands indicate a variation of $\eta\in[1.8,2.0]$.
                    The experimental data for $G_E^n$ and $G_M^n$ are from Refs.~\cite{:2008ha,Passchier:1999cj,Herberg:1999ud,Zhu:2001md,Bermuth:2003qh,Warren:2003ma,Glazier:2004ny,Plaster:2005cx,Riordan:2010id}
                    and~\cite{Rock:1982gf,Lung:1992bu,Anklin:1998ae,Kubon:2001rj,Anderson:2006jp,Lachniet:2008qf}, respectively.
                    The references for $G_E^p$ are given in Fig.~\ref{fig:ratios} and those for $G_M^p$ can be found in~\cite{Perdrisat:2006hj}.} \label{fig:ffs}

                    \end{center}
        \end{figure*}

    \renewcommand{\arraystretch}{1.0}

             \begin{table*}[t]

                \begin{center}
                \begin{tabular}{   c @{\;\;} ||  @{\;\;}c@{\;\;} || @{\;\;}c@{\;\;} | @{\;\;}c@{\;\;} | @{\;\;}c@{\;\;} | @{\;\;}c@{\;\;} ||  @{\;\;}c@{\;\;} | @{\;\;}c@{\;\;} | @{\;\;}c@{\;\;} | @{\;\;}c@{\;\;}    }

                                    &  $M_N$         &     $\mu^p$   &  $\mu^n$     &  $\kappa^v$  &  $\kappa^s$ &  $r_E^p$   &  $(r_E^n)^2$ &  $r_M^p$ &  $r_M^n$        \\   \hline

                    Faddeev         &  $0.94$        &     $2.21(1)$ &   $-1.33(1)$ &  $2.54(2)$   &  $-0.12(1)$ &  $0.75(3)$ &  $-0.01$     &  $0.72(2)$ &  $0.72(2)$               \\
                    Exp.            &  $0.94$        &     $2.79$   &   $-1.91$     &  $3.70$      &  $-0.12$    &  $0.89$    &  $-0.12$     &  $0.86$    &  $0.87$

                \end{tabular} \caption{Results for the nucleon mass and static electromagnetic properties compared to experiment.
                                       The parentheses indicate the dependence on the infrared parameter $\eta$ of Eq.~\eqref{couplingMT}.
                                       $M_N$ is given in GeV, the magnetic moments are expressed in nuclear magnetons, and the charge radii are given in fm.
                                       }\label{tab:results}
                \end{center}

        \end{table*}

\section{Results and discussion} \label{sec:results}

       We computed the electromagnetic form factors of the nucleon within the decomposition of Fig.~\ref{fig:faddeev-current-specific} for the electromagnetic current.
            We emphasize that the present setup allows to determine \textit{all} ingredients of the electromagnetic current
            without any further approximations:
             the rainbow-ladder kernel that appears in its definition
            is given in Eq.~\eqref{RLkernel}; the dressed quark propagator is obtained from its Dyson-Schwinger equation~\eqref{quarkdse};
            the nucleon amplitude is the solution of the covariant Faddeev equation \eqref{faddeev:eq}; and the quark-photon vertex is obtained consistently from its
            inhomogeneous Bethe-Salpeter equation, also solved within a rainbow-ladder truncation, and is described in App.\,\ref{app:qpv}.
            Apart from the current-quark mass, the single parameter that enters the equations and controls the results is the infrared scale $\Lambda$
            which appears in Eq.~\eqref{couplingMT} in connection with the rainbow-ladder truncation, where it was fixed to reproduce the experimental pion decay constant.
            In the chiral limit, $\Lambda$ is the only relevant scale in the system, and all mass-dimensionful quantities scale with $\Lambda$.
            This entails that dimensionless form factors become independent of this scale.
            One can further investigate the impact of the infrared properties, described by the width parameter $\eta$ (cf.~Fig.~\ref{fig:coupling}),
            on resulting observables. This is indicated by the bands in Figs.~(\ref{fig:ffs}--\ref{fig:ratios}), and in the same way as $\pi-$ and $\rho-$
            ground state properties are not very sensitive to $\eta$, this sensitivity is found to be weak in the case of nucleon form factors as well.

       \subsection{Low-momentum behavior}

       Fig.~\ref{fig:ffs} shows the results for the nucleon electromagnetic form factors,
       calculated at the physical $u/d$ point corresponding to $m_\pi=140$~MeV and compared to experiment.
       We find a remarkable agreement with the experimental data above a photon momentum transfer $Q^2 \gtrsim 2$ GeV$^2$.
       This is the region where pseudoscalar-meson cloud effects should vanish as the photon probes the nucleon at length scales
       much smaller than the typical size of pionic correlations 
       and thereby essentially reveals the nucleon's quark core.
       An illustrative example is the neutron's electric form factor where our curve is close to the new Hall-A data from JLAB~\cite{Riordan:2010id},
       whereas the bump at low $Q^2$ is completely absent in our result.
       This suggests that the low-$Q^2$ structure of $G_E^n$ predominantly owes
       to virtual pion-cloud components.
       Similar effects are found in the proton's and neutron's magnetic form factors: below $Q^2\sim 2$ GeV$^2$, the
       results underestimate the data and, at vanishing photon momentum, yield magnetic moments
       that are $20-30\%$ smaller than the experimental values, see Table~\ref{tab:results}: 
             for the proton and neutron, they read $\mu^p=2.21(1)$ and $\mu^n=-1.33(1)$, where the brackets denote the sensitivity to $\eta$.

              \begin{figure*}[htp]
                         \begin{center}

                         \includegraphics[scale=0.33]{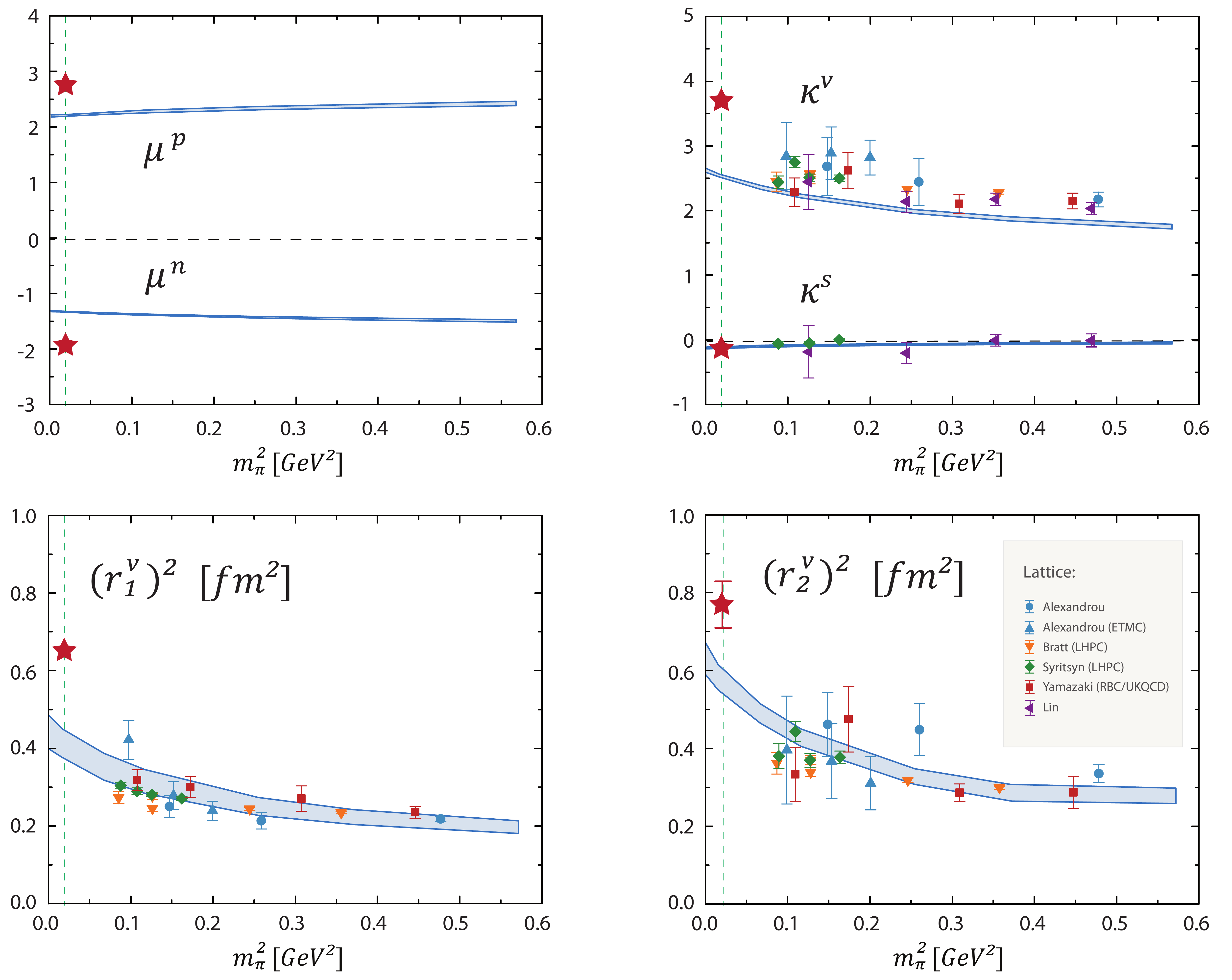}
                         \caption{(Color online) Static electromagnetic properties of the nucleon as a function of the squared pion mass, for $\eta = 1.8\pm 0.2$.
                                                 \textit{Upper left panel:} dimensionless magnetic moments of proton and neutron.
                                                 \textit{Upper right panel:} isovector and isoscalar magnetic moments in units of static nuclear magnetons.
                                                 \textit{Lower panels:} squared isovector Dirac and Pauli radii, given in fm$^2$.
                                                 The lattice data are from Refs.~\cite{Alexandrou:2006ru,Alexandrou:2008rp,Syritsyn:2009mx,Yamazaki:2009zq,Bratt:2010jn,Lin:2010fv}.
                                                 Stars denote experimental values~\cite{Belushkin:2006qa,Nakamura:2010zzi} at the $u/d-$quark mass
                                                 whose position is indicated by dashed vertical lines.
                                                 }\label{fig:static}

                         \end{center}
             \label{fig.2}
             \end{figure*}

             Fig.~\ref{fig:static} displays the current-quark mass evolution of our calculated nucleon magnetic moments and charge radii,
             again expressed in terms of the squared pion mass which is obtained from its Bethe-Salpeter equation.
             The upper left panel depicts the dimensionless magnetic moments $\mu^{p,n}=G_M^{p,n}(0)$
             which exhibit a moderately rising behavior with the pion mass.
             In the upper right panel of Fig.~\ref{fig:static} the quark-mass dependence
             of the isovector and isoscalar anomalous magnetic moments $\kappa^v=\kappa^p-\kappa^n$
             and $\kappa^s=\kappa^p+\kappa^n$, expressed in static nuclear magnetons, is compared to recent lattice data.
             A very similar current-mass dependence emerges in the quark-diquark model~\cite{Eichmann:2010je};
             however, compared to those results, the gluon-interaction kernel in the explicit three-quark framework
             leads to an overall reduction of $10-15\%$ for the magnetic moments.

             For a comparison of magnetic moments in units of static nuclear magnetons with experiment or lattice,
             one must bear in mind that our calculated nucleon mass differs (although marginally, cf. Fig.~\ref{fig:nucleonmass}) from
             that obtained on the lattice at higher pion masses.
             The nucleon's magnetic moments are given by
             \begin{equation}\label{magneticmoment}
                 \mu^{p,n}_\text{dim} = \frac{e}{2M}\,G^{p,n}_{M}(0) = \frac{e}{2M_\text{exp}}\left[ G^{p,n}_{M}(0)\,\frac{M_\text{exp}}{M}\right],
             \end{equation}
             where $M$ is running with the current-quark mass; hence their values in static nuclear magnetons are given by
             the bracket in Eq.\,\eqref{magneticmoment}.
             The unambiguous comparison of magnetic moments is that of the dimensionless value $G_{M}(0)$, whereas
             the corresponding magnetic moments in static nuclear magnetons will compare differently if the nucleon masses in both approaches do not coincide.
             To account for this, we plot $\kappa^{v,s}$ in Fig.~\ref{fig:static} by replacing $M$ in Eq.~\eqref{magneticmoment} with
             the following reference mass~\cite{Eichmann:2010je}:
                  \begin{equation}\label{referencemass}
                      M^2_\text{Ref}(m_\pi^2) = M_0^2 + \left(\frac{3 m_\pi}{2}\right)^2 \left(1 + f(m_\pi^2)\right) ,
                  \end{equation}
             where $f(m_\pi^2) = 0.77/(1+(m_\pi/0.65\,\text{GeV})^4)$ and $M_0=0.9$~GeV.
             Eq.\,\eqref{referencemass} reproduces the experimental nucleon mass at $m_\pi=0.14$ GeV, approaches the heavy-quark limit via $M \rightarrow 3m_\pi/2$ and
             describes the dynamical lattice results for the nucleon mass in Fig.~\ref{fig:nucleonmass} reasonably well.

              Chiral cloud corrections to core magnetic moments can be estimated
              from the pion-loop contributions $\delta\mu^{p,n} = \mu^{p,n} - \mu^{p,n}_\text{core}$  in heavy-baryon chiral effective field theory.
              In that framework
              the sum of the two diagrams where the photon strikes a pion in the nucleon and leaves an intermediate
              nucleon or $\Delta-$baryon as a spectator is given by~\cite{Young:2004tb,Wang:2007iw}
              \begin{equation}\label{mm-pionloops}
                  \delta\mu^{p,n} = \pm \lambda \int\limits_0^\infty dx\, \frac{x^4}{\omega^4}\,u(x)^2
                                                       \left[ 1 + \frac{C^2}{9\,g_A^2}\,\frac{\omega\,(2\omega+1)}{(\omega+1)^2}\right],
              \end{equation}
              where $x=|\vect{k}|/(\Delta M)$ and $\omega=\sqrt{x^2+m_\pi^2/(\Delta M)^2}$ are the pion momentum
              and intermediate pion energy, normalized by the physical $N$--$\Delta$ mass splitting $\Delta M=0.29$~GeV.
              The constants are given by
              $C=-2\!\cdot\!(0.76)$ and $\lambda= g_A^2 M_N (\Delta M) /(3\pi^2 f_\pi^2)$ with $f_\pi=131$ MeV and $g_A=1.26$.
              Choosing an identical dipole ansatz $u(x) = [ 1 + (|\vect{k}|/\Lambda)^2 ]^{-2}$ for the $NN\pi$ and $N\Delta\pi$ vertex dressings
              with a regulator $\Lambda=0.8$ GeV yields the pion-loop contribution $\delta\mu^{p,n} = \pm 0.61$ at the physical pion mass; however,
              the inclusion of further meson-loop diagrams can diminish this value~\cite{Wang:2007iw}.
              Upon subtracting this result from the experimental values one obtains the core estimates $\mu^p_\text{core}=2.18$ and $\mu^n_\text{core}=-1.30$.
              This is remarkably close to our calculated results $\mu^p = 2.21(1)$ and $\mu^n = -1.33(1)$.

              Since the simplest pion-loop contributions to proton and neutron magnetic moments of Eq.~\eqref{mm-pionloops} carry an opposite sign, their
              total cancels in the isoscalar combination.
              This implies $\kappa^s \approx \kappa^s_\text{core}$, i.e. isoscalar magnetic moments should be roughly undisturbed by pion-cloud corrections.
              The experimental isoscalar anomalous magnetic moment is small and negative, $\kappa^s=-0.12$,
              and reproduced by the Faddeev calculation, cf. Table~\ref{tab:results}, with only a small model dependence.

        The electromagnetic radius corresponding to a form factor $G_i(Q^2)$ is defined as
        \begin{equation}
           r_i^2 = -\frac{6}{G_i(0)} \left.\frac{dG_i}{dQ^2}\right|_{Q^2=0}\,(\hbar c)^2\,,
        \end{equation}
        where $\hbar c =0.197$ GeV fm, and $G_i(0)$ in the denominator is dropped for the neutron's form factors $G_E^n$ and $F_1^n$   
        as they vanish at $Q^2=0$.
        The lower panels in Fig.~\ref{fig:static} show the squared isovector Dirac and Pauli radii $(r_1^v)^2$ and $(r_2^v)^2$,
        compared to lattice data. Once again, the unambiguous comparison with the lattice is that
        of the dimensionless values $r_i^2 M^2$, hence we rescale our calculated radii via
             \begin{equation}\label{radii-rescaled}
                 r^2_\text{resc} = r^2_\text{calc}\left(\frac{M_\text{calc}}{M_\text{Ref}}\right)^2\,,
             \end{equation}
        with $M_\text{Ref}$ from Eq.~\eqref{referencemass}.
        The plot shows a satisfactory agreement with the lattice data at larger pion masses
        where the pion-cloud dressing effects of the nucleon are diminished. 
        Compared to the reduction of the magnetic moments, the isovector radii from the Faddeev approach are virtually identical to those obtained in the quark-diquark model~\cite{Eichmann:2010je}.
        The absence of a meson cloud is further signaled by missing chiral curvature 
        as the charge radii of the nucleon, surrounded by a pion cloud, would diverge in the chiral limit.
        At the $u/d$ mass, the experimental radii $r_E^p$, $r_M^p$ and $r_M^n$ are underestimated by $15-20\%$, cf. Table~\ref{tab:results}.

        To summarize this section: the combined behavior of the nucleon's magnetic moments and the electromagnetic radii,
        compared to experiment and lattice data,
        provides strong evidence that the main missing ingredients to nucleon form factors in a rainbow-ladder calculation
        are those induced by chiral-cloud corrections. Thus, Fig.~\ref{fig:static} outlines the characteristic features
        of a current-mass dependent nucleon quark core.

       \begin{figure*}[tp]
                    \begin{center}

                    \includegraphics[scale=0.37]{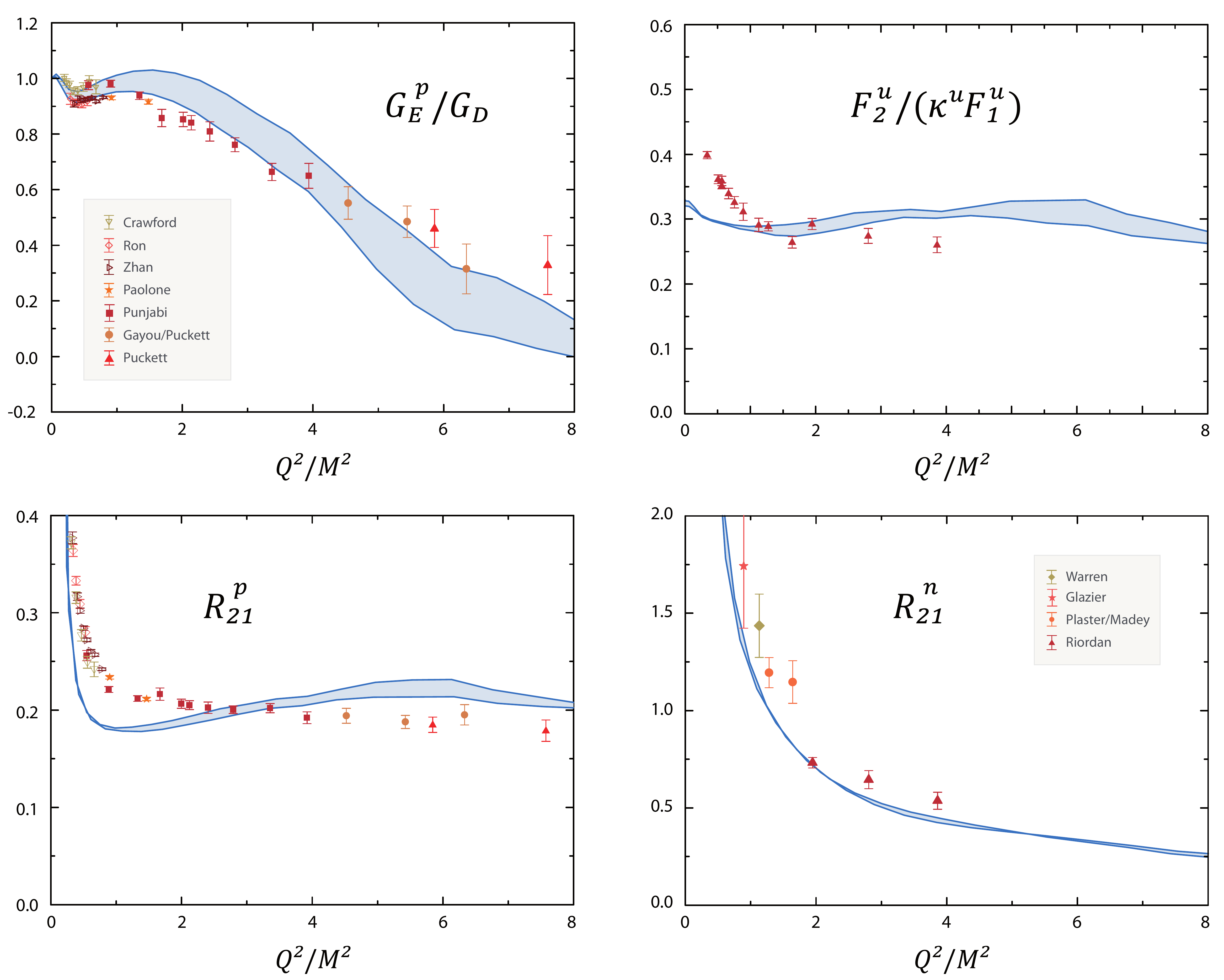}
                    \caption{(Color online) $Q^2-$evolution of nucleon electromagnetic form factor ratios. The bands denote the $\eta$ variation of Fig.~\ref{fig:ffs}.
                    \textit{Upper left panel:} Electric proton form factor normalized by the dipole (see text) and compared to experimental data
                    from Refs.~\cite{Punjabi:2005wq,Crawford:2006rz,Paolone:2010qc,Puckett:2010ac,Ron:2011rd,Zhan:2011ji,Puckett:2011xg}.
                    \textit{Upper right panel:} Pauli-to-Dirac ratio for the up-quark, normalized by the experimental value $\kappa^u=1.67$, with data from Ref.~\cite{Cates:2011pz}.
                    \textit{Lower panels:} weighted Pauli-to-Dirac ratios of Eq.~\eqref{weighted-Pauli-to-Dirac-ratio} for proton and neutron.
                    The experimental data for $R_{21}^p$ are the same as in the upper left panel and those for $R_{21}^n$ are identical to $G_E^n$ in Fig.~\ref{fig:ffs}.   } \label{fig:ratios}

                    \end{center}
        \end{figure*}

       \subsection{Large-momentum behavior}

       With the body of new experimental data at higher photon momentum transfer,    
       the large-$Q^2$ behavior of the nucleon's electromagnetic form factors has attracted considerable attention
       in the past decade.
              Perturbative QCD predicts the scaling behavior of the Dirac
              and Pauli form factors as~\cite{Brodsky:1974vy}                                                                
              \begin{equation}
                  F_1 \rightarrow 1/Q^4\,, \quad F_2 \rightarrow 1/Q^6\,, \quad Q^2 \,F_2/F_1 \rightarrow const.
              \end{equation}
              Correspondingly, the Sachs form factors would scale as $G_E, G_M \rightarrow 1/Q^4$ which implies
              that the ratio $G_E/G_M$ becomes constant.
              These predictions have come under scrutiny with the polarization-transfer measurements, 
              where the ratio $G_E^p/G_M^p$ shows roughly a linear decrease with $Q^2$ and points toward a zero crossing at some larger value of $Q^2$.

              The discrepancy between the perturbative prediction and the experimental data has been attributed to
              helicity non-conservation and the presence of
              non-zero quark orbital angular-momentum in the nucleon amplitude~\cite{Miller:2002qb,Ralston:2003mt,Holl:2005zi}.    
              Indeed, with an updated perturbative prediction for $F_2$ that accounts for wave-function components with orbital angular momentum~\cite{Belitsky:2002kj},
             \begin{equation} \label{F2-F1-perturbative}
                 Q^2\,(F_2/F_1)/\ln^2{ \left(Q^2/\Lambda^2\right)} \rightarrow const.,
             \end{equation}
              the onset of perturbative scaling in the proton's Pauli to Dirac ratio appears to happen already at comparatively low photon momenta, cf.~Fig.~\ref{fig:ratios}.

       While the large-$Q^2$ behavior is in principle accessible in the Faddeev approach, its current implementation
       is limited by an upper value of the photon momentum. The quark propagator obtained from the DSE
       inevitably develops a certain singularity structure which can be characterized by a pole mass $m_q$~\cite{Eichmann:2009zx}.
       The photon probes the quark propagator $S(p)$ within a parabola in the complex $p^2-$domain
       that grows with $Q^2$ and is thereby limited by the value of $m_q$.
       In absence of a sophisticated method to account for these singularities, a bound-state amplitude
       can be computed from the Faddeev equation only if $M<3m_q$, and the form-factor calculation is restricted to $Q^2<4\left((3m_q)^2-M^2\right)\approx 7\dots 9$~GeV$^2$,
       depending on the infrared parameter $\eta$. In proximity to this upper bound the results become sensitive to the numerics
       and require an increasingly better resolution of the Faddeev amplitude.
       Thus, with presently available accuracy, the results in this region should be interpreted with caution.

       The upper left panel of Fig.~\ref{fig:ratios} shows the electric form factor of the proton normalized by the dipole ansatz $G_D = (1+c_D \,Q^2/M^2)^{-2}$,
       where $c_D = M_\text{exp}^2/(0.71\,\text{GeV}^2)$. It exhibits a fall-off in $Q^2$ which
       signals the presence of orbital angular momentum in the Faddeev amplitude, cf.~Table~\ref{tab:spd}.
       While we do not yet deem our numerical results fully reliable in the large-momentum region, Fig.~\ref{fig:ratios}
       demonstrates that a zero crossing can occur quite naturally within the Faddeev approach.
       In fact, if Eq.~\eqref{F2-F1-perturbative} is valid at already moderate photon momenta
       it easily accommodates a zero crossing for $G_E^p/G_M^p$; namely,
       through Eq.~\eqref{ff:sachs}, at $\ln^2(Q^2/\Lambda^2)=4M^2/const.$
       Quark-diquark model studies typically find          
       a zero crossing as well, where its location depends on the model parameters in the calculation~\cite{Holl:2005zi,Cloet:2008re}.

       In connection with this, an essential component to ensure the $Q^2$-evolution in Figs.~\eqref{fig:ffs} and~\eqref{fig:ratios}
       is the quark-photon vertex solution from its inhomogeneous BSE.
       While the Ball-Chiu ansatz of Eq.~\eqref{vertex:BC} alone is sufficient to guarantee current conservation,
       it misses important transverse parts in the vertex which are dominated by a vector-meson pole in the vertex at $Q^2=-m_\rho^2$~\cite{Maris:1999bh}.
       It has been noted in several works that the Ball-Chiu vertex contributes only $\sim 50\%$ to squared hadronic charge radii~\cite{Maris:1999bh,Bhagwat:2006pu,Eichmann:2008ae,Eichmann:2010je}.
       That statement stays true here.
       While the magnetic moments are not very sensitive to these transverse contributions,
       the $Q^2-$evolution changes dramatically:
       at $Q^2/M^2=3$, the form factors $G_E^n$, $G_M^p$ and $G_M^n$ obtained from the Ball-Chiu ansatz alone overestimate the results in Fig.~\ref{fig:ffs}
       by a factor of 3, whereas $G_E^p$ develops a zero crossing at $Q^2/M^2\approx 4$.
       This makes clear that the position of a zero crossing might well depend on the truncation,
       as effects beyond rainbow-ladder could certainly impact upon the transverse structures of the quark-photon vertex
       and thereby play a delicate role in the form factors at larger $Q^2$.

       The lower panels of Fig.~\ref{fig:ratios} display the weighted ratio of Pauli to Dirac form factors for proton and neutron,
       \begin{equation}\label{weighted-Pauli-to-Dirac-ratio}
           R_{21} := \frac{Q^2}{M^2}\,\frac{F_2}{F_1}\,\frac{1}{\ln^2{ \left(c_R \,Q^2/M^2\right)}}\,,
       \end{equation}
       where $c_R = (M_\text{exp}/(0.3\,\text{GeV}))^2$.
       Eq.~\eqref{F2-F1-perturbative} entails that the ratios $R_{21}^{p,n}$ should approach constant values at large photon momenta.
       While such a near constancy in $R_{21}^p$ is observed already at moderate $Q^2$, the neutron's ratio does not yet follow this trend.
       Irrespective of that, we find a fair agreement between our results and the data.

       An intriguing feature which was recently noted~\cite{Cates:2011pz} is the behavior of the up- and down-quark ratios
       $F_2^u/F_1^u$ and $F_2^d/F_1^d$ above $Q^2\sim 1$ GeV$^2$: compared to their proton and neutron counterparts, they are
       roughly constant, whereas $F_2^u/F_1^u$ rises sharply below that value toward its static value $\nicefrac{1}{2}$, as shown in the upper right panel of Fig.~\ref{fig:ratios}.
       Combined with the suppression of the $d-$quark form factors compared to those of the $u-$quark this explains the observed behavior of $R_{21}^p$ and $R_{21}^n$.
       The decline of $F_1^d/F_1^u$ and $F_2^d/F_2^u$
       is responsible for the negative form factors $F_{1,2}^n$ of the neutron.
       In the infinite-momentum frame it is related to a
       concentration 
       of the up-quark distribution in the proton, and that of the $d-$quark in the neutron, in impact-parameter space~\cite{Miller:2008jc}.
       In the quark-diquark model this suppression owes to the singly represented $d-$quark in the proton
       that mainly interacts with the photon in combination with an axial-vector diquark in a $uu$ configuration~\cite{Cloet:2008re}.
       The upper right panel of Fig.~\ref{fig:ratios} shows that the constancy of $F_2^u/F_1^u$ also maintains for higher values of the photon momentum,
       whereas the rise at small $Q^2$ is practically absent and thereby likely to be related to pion-cloud effects.

        We finally remark that there is no qualitative difference in the form factors' $Q^2-$evolution
        up to the strange-quark mass: both the form factor curves of Fig.~\ref{fig:ffs}
        as well as their ratios in Fig.~\ref{fig:ratios} essentially retain their shape
        when plotted as a function of the dimensionless variable $Q^2/M^2$.
        This feature was also observed in Ref.\,\cite{Nicmorus:2010sd} for the $\Delta$ electromagnetic form factors in the quark-diquark context.
        It can be understood from the correlations in the Faddeev amplitude, cf. Table~\ref{tab:spd}:
        between the up-/down and strange-quark regime, orbital angular momentum contributes one-third to the nucleon spin
        and this contribution only slowly decreases with rising current-quark mass.
        For the quark masses investigated here,
        the disappearance of $p-$wave components in the nucleon's rest frame which is expected to have
        a sizeable impact on nucleon form factors is therefore not yet visible in our results.

\section{Conclusions}

      We presented a calculation of nucleon electromagnetic form factors in the Poincar\'e-covariant Faddeev approach,
      based on the Dyson-Schwinger equations of QCD.
      We employed a rainbow-ladder truncation which explains the binding in the nucleon through iterated dressed gluon exchange between the quarks.
      Thereby all ingredients of the equations are completely fixed and no further model assumptions need to be made.
      The single parameter of the approach is a scale which is fixed to reproduce the pion decay constant.
      The nucleon's electromagnetic current in terms of three interacting quarks was derived.
      Since pion-cloud effects are not implemented, our results represent the nucleon's quark core.

      We find a good agreement between our form-factor results and the experimental data above $Q^2 \approx 2$ GeV$^2$.
      The form-factor ratio $G_E^p/G_M^p$ falls off with larger $Q^2$ and the results for the Pauli to Dirac ratios closely follow the data.
      These features can be attributed to a significant amount of orbital angular momentum which appears in the solution for the nucleon's amplitude
      and persists up to the strange quark mass, and most likely much farther beyond.
      In the low-$Q^2$ region clear signals of missing pion-cloud effects are present:
      the charge radii and magnetic moments underestimate the data, and the enhanced low-$Q^2$ structure in the neutron's electric form factor is absent.
      At larger quark masses, where pionic effects do no longer contribute,
      our results for the nucleon magnetic moments and electric charge radii are comparable to those obtained in lattice QCD.

      The present framework does not include genuine three-quark interactions.
      However, our results suggest that such contributions are indeed small         
      and that it is essentially the quark-quark correlations which contribute most of the nucleon's binding.
      The overall agreement between our results and those obtained in the quark-diquark model provides further evidence
      for the quark-diquark structure of the nucleon,
      and it implies that scalar and axial-vector diquark degrees of freedom
      can account for most of its characteristic features.

      In view of a better understanding of the chiral and low-$Q^2$ structure of baryons,
      our approach must be improved by incorporating missing chiral cloud effects.
      This is a challenging task as it necessitates a consistent description of baryons beyond the rainbow-ladder truncation.
      Steps in that direction are planned.
      Moreover, an investigation of nucleon electromagnetic form factors at large photon momentum as well as nucleon-to-resonance transition form factors is desirable.
      The results presented herein are encouraging, and
      we are confident that the covariant Faddeev framework provides a suitable starting point for future studies of baryon structure.


     \section{Acknowledgements}

     I am grateful to R. Alkofer, M. Blank, I. C. Clo\"et, C. S. Fischer, A. Krassnigg, D. Nicmorus, H. Sanchis-Alepuz and R. Williams for valuable discussions.
     This work was supported by the Austrian Science Fund FWF under
     Erwin-Schr\"odinger-Stipendium No.~J3039 and
     the Helmholtz International Center for FAIR 
     within the framework of the LOEWE program launched by the
     State of Hesse, GSI, BMBF and DESY.


\begin{appendix}

  \section{Conventions and formulas} \label{app:conventions}

        \subsection{Euclidean conventions}

            We work in Euclidean momentum space with the following conventions:
            \begin{equation}
                p\cdot q = \sum_{k=1}^4 p_k \, q_k,\quad
                p^2 = p\cdot p,\quad
                \Slash{p} = p\cdot\gamma\,.
            \end{equation}
            A vector $p$ is spacelike if $p^2 > 0$ and timelike if $p^2<0$.
            The hermitian $\gamma-$matrices $\gamma^\mu = (\gamma^\mu)^\dag$ satisfy the anticommutation relations
            $\left\{ \gamma^\mu, \gamma^\nu \right\} = 2\,\delta^{\,\mu\nu}$, and we define
            \begin{equation}
                \sigma^{\mu\nu} = -\frac{i}{2} \left[ \gamma^\mu, \gamma^\nu \right]\,, \quad
                \gamma^5 = -\gamma^1 \gamma^2 \gamma^3 \gamma^4\,.
            \end{equation}
            In the standard representation one has:
            \begin{equation*}
                \gamma^k  =  \left( \begin{array}{cc} 0 & -i \sigma^k \\ i \sigma^k & 0 \end{array} \right), \;
                \gamma^4  =  \left( \begin{array}{c@{\quad}c} \mathds{1} & 0 \\ 0 & \!\!-\mathds{1} \end{array} \right), \;
                \gamma^5  =  \left( \begin{array}{c@{\quad}c} 0 & \mathds{1} \\ \mathds{1} & 0 \end{array} \right),
            \end{equation*}
            where $\sigma^k$ are the three Pauli matrices.
            The charge conjugation matrix is given by
            \begin{equation}
                C = \gamma^4 \gamma^2, \quad C^T = C^\dag = C^{-1} = -C\,,
            \end{equation}
            and the charge conjugates for (pseudo-)\,scalar, \mbox{(axial-)} vector and tensor amplitudes are defined as
            \begin{equation}\label{chargeconjugation}
            \begin{split}
                \conjg{\Gamma}(p,P) &:= C\,\Gamma(-p,-P)^T\,C^T \,,   \\
                \conjg{\Gamma}^\alpha(p,P) &:= -C\,{\Gamma^\alpha}(-p,-P)^T\,C^T \,,   \\
                \conjg{\Gamma}^{\beta\alpha}(p,P) &:= C\,{\Gamma^{\alpha\beta}}(-p,-P)^T\,C^T\,,
            \end{split}
            \end{equation}
            where $T$ denotes a Dirac transpose.
            Four-momenta are conveniently expressed through hyperspherical coordinates:
            \begin{equation}\label{APP:momentum-coordinates}
                p^\mu = \sqrt{p^2} \left( \begin{array}{l} \sqrt{1-z^2}\,\sqrt{1-y^2}\,\sin{\phi} \\
                                                           \sqrt{1-z^2}\,\sqrt{1-y^2}\,\cos{\phi} \\
                                                           \sqrt{1-z^2}\;\;y \\
                                                           \;\; z
                                         \end{array}\right),
            \end{equation}
            and a four-momentum integration reads:
            \begin{equation*} \label{hypersphericalintegral}
                 \int\limits_p := \frac{1}{(2\pi)^4}\,\frac{1}{2}\int\limits_0^{\infty} dp^2 \,p^2 \int\limits_{-1}^1 dz\,\sqrt{1-z^2}  \int\limits_{-1}^1 dy \int\limits_0^{2\pi} d\phi \,.
            \end{equation*}

        \subsection{Transverse identities}

 \renewcommand{\arraystretch}{1.3}

            The Faddeev-equation formalism simplifies when working with quantities that are transverse to the nucleon momentum $P$, where $P^2 = -M^2$.
            Covariant relations between transverse objects bear a close resemblance to those in three-dimensional Euclidean space:
            in the nucleon's rest frame one has $\hat{P}=e_4$, and quantities transverse to $e_4$ reduce to the usual three-dimensional form.
            In a form factor calculation, typically carried out in the Breit frame, the average momentum $(P_i+P_f)/2$ of incoming and outgoing
            nucleon is at rest.
            We define the transverse projector with respect to $P$:
            \begin{equation}
               T_P^{\mu\nu} = \delta^{\mu\nu} - \hat{P}^\mu \hat{P}^\nu \,,
            \end{equation}
            as well as $\gamma-$matrices transverse to $P$:
            \begin{equation}
                \gamma_T^\mu = T_P^{\mu\nu} \gamma^\nu = \gamma^\mu - \hat{P}^\mu \hat{\Slash{P}} \,,
            \end{equation}
            which satisfy $\left\{ \gamma_T^\mu, \gamma_T^\nu \right\} = 2\,T_P^{\mu\nu}$ and $\gamma_T^\mu\, \gamma_T^\mu =3$.
            Then $\gamma_T^\mu$, $\gamma_5$ and $\hat{\Slash{P}}$ pairwise anticommute.
            Similarly, we denote momenta transverse to $P$ by
            \begin{equation}
                p_T^\mu = T_P^{\mu\nu} p^\nu = p^\mu - p\cdot\hat{P}\,\hat{P}^\mu
            \end{equation}
            and define
            \begin{equation}
                \sigma_T^{\mu\nu} = -\frac{i}{2}\,[\gamma_T^\mu,\gamma_T^\nu] = i (T_P^{\mu\nu} - \gamma_T^\mu \gamma_T^\nu) \,.
            \end{equation}
            We make frequent use of the positive- and negative energy projectors
            \begin{equation}\label{energy-projectors}
               \Lambda_\omega(P) = (\mathds{1} +\omega \hat{\Slash{P}} )/2\,, \quad \omega=\pm
            \end{equation}
            which obey the relations
            \begin{equation}\label{energy-projectors-properties}
                \hat{\Slash{P}}\,\Lambda_\omega = \Lambda_\omega\,\hat{\Slash{P}} = \omega\,\Lambda_\omega\,, \quad
                \begin{array}{rl}
                \Lambda_\omega\,\gamma_5 &= \gamma_5\,\Lambda_{-\omega}\,, \\
                \Lambda_\omega\,\gamma_T^\mu &= \gamma_T^\mu\,\Lambda_{-\omega}\,.
                \end{array}
            \end{equation}
            It is convenient to define a transverse $\epsilon-$tensor by
            \begin{equation}\label{transverse-epsilon}
                \epsilon^{\alpha\beta\gamma}_T := \epsilon^{\alpha\beta\gamma\lambda}\,\hat{P}^\lambda\,,
            \end{equation}
            where $\epsilon^{1234} = 1$,
            which satisfies
            \begin{equation}
            \begin{split}
                \epsilon^{\mu\alpha\beta}_T\,\epsilon^{\mu\rho\sigma}_T &= T_P^{\alpha\rho}\,T_P^{\beta\sigma}-T_P^{\alpha\sigma}\,T_P^{\beta\rho}\,, \\
                \epsilon^{\mu\nu\alpha}_T\,\epsilon^{\mu\nu\beta}_T &= 2 \,T_P^{\alpha\beta}\,, \\
                \epsilon^{\mu\nu\rho}_T\,\epsilon^{\mu\nu\rho}_T &= 6
            \end{split}
            \end{equation}
            and
            \begin{align}
                 \gamma_5\,\epsilon^{\alpha\beta\gamma}_T \hat{\Slash{P}} &= T^{\alpha\beta}_P \gamma^\gamma_T + T^{\beta\gamma}_P \gamma^\alpha_T - T^{\gamma\alpha}_P \gamma^\beta_T \qquad \nonumber \\
                                                                                                &- \gamma_T^\alpha\,\gamma_T^\beta\,\gamma_T^\gamma\,, \\
                 \gamma_T^\mu \,\gamma_5\,\epsilon^{\mu\alpha\beta}_T  \hat{\Slash{P}} &= T_P^{\alpha\beta} - \gamma_T^\alpha \,\gamma_T^\beta \,, \label{epsilon-id-2}\\
                 \gamma_T^\mu \,\gamma_T^\nu \,\gamma_5 \,\epsilon^{\mu\nu\alpha}_T  \hat{\Slash{P}} &= 2 \gamma_T^\alpha\,, \\
                 \gamma_T^\mu \,\gamma_T^\nu \,\gamma_T^\rho \,\gamma_5 \,\epsilon^{\mu\nu\rho}_T  \hat{\Slash{P}} &= 6\,.
            \end{align}

  \section{Nucleon amplitude} \label{sec:nucamp}

 \renewcommand{\arraystretch}{1.6}

             The nucleon's covariant three-quark amplitude including its full Dirac, flavor and color dependence is given by
              \begin{equation}\label{FE:nucleon_amplitude_full}
                  \mathbf{\Psi}(p,q,P) = \left( \sum_{n=1}^2  \Psi_n \, \mathsf{F}_n\right) \frac{\varepsilon_{ABC}}{\sqrt{6}}\,,
              \end{equation}
             where the antisymmetric color part is normalized to $1$.
             $(\Psi_n)_{\alpha\beta\gamma\delta}(p,q,P)$ and $(\mathsf{F}_n)_{abcd}$
             are the spin-momentum and flavor amplitudes which transform as doublets under
             the permutation group $\mathds{S}^3$, with entries for $n=1,2$.
            The Dirac amplitudes $\Psi_n$ carry 3 spinor indices for the quark legs $(\alpha,\beta,\gamma)$ and one
            spinor index $\delta$ for the nucleon. They are mixed-antisymmetric $(\Psi_1)$ or mixed-symmetric $(\Psi_2)$
            under exchange of the indices $\alpha$, $\beta$ and corresponding quark momenta.
            Likewise, the two isospin-$\nicefrac{1}{2}$ flavor tensors $\mathsf{F}_n$ carry 3 isospin indices
            $(a,b,c)$ for the quarks and one ($d$) for the nucleon and are mixed-(anti-)symmetric in $a$, $b$. They read:
              \begin{equation} \label{FAD:flavor}
              \begin{split}
                  \mathsf{F}_1 &= \textstyle\frac{1}{\sqrt{2}}\,i \sigma_2\otimes \mathds{1} \,,\\
                  \mathsf{F}_2 &= -\textstyle\frac{1}{\sqrt{6}}\, \vect{\sigma}\,i\sigma_2 \otimes \vect{\sigma} \,,
              \end{split}
              \end{equation}
              where $\sigma_i$ are the Pauli matrices and
            the tensor product is understood as
            \begin{equation}
            \begin{split}
                (f\otimes g)_{abcd} &= f_{ab} \,g_{cd}\,, \\
                (f_1\otimes f_2)(g_1\otimes g_2) &= (f_1\,g_1)\otimes(f_2\, g_2)\,.
            \end{split}
            \end{equation}
             They are normalized to $(\mathsf{F}^\dag_{n'})_{bad'c}(\mathsf{F}_n)_{abcd} = \delta_{nn'} \delta_{dd'}$.
             To project onto proton or neutron flavor states, the index $d$ is contracted with either of the two isospin vectors $(1,0)$ or $(0,1)$.

        \subsection{Permutation-group properties}\label{sec:permutationgroup}

 \renewcommand{\arraystretch}{1.1}

             The Pauli principle requires the full Dirac-flavor-color amplitude of Eq.~\eqref{FE:nucleon_amplitude_full} to be antisymmetric
             under exchange of any two quark legs, i.e. under a combined permutation of Dirac, flavor and color indices and
             the corresponding exchange of momenta. The color amplitude in Eq.~\eqref{FE:nucleon_amplitude_full} is an antisymmetric singlet.
             The transformation properties of the doublets~\eqref{FAD:flavor} follow from Fierz identities. With the constraint
             of a fully symmetric Dirac-flavor part one thus obtains the transformation behavior of the Dirac amplitudes.
             We abbreviate the 6 possible permutations by 
 \renewcommand{\arraystretch}{1.5}
             \begin{equation*}
             \begin{array}{l @{\quad} l @{\,} l @{\qquad} l @{\,} l}
                  (123):  & \mathsf{F}_n &= (\mathsf{F}_n)_{abcd}\,,  & \Psi_n  &= (\Psi_n)_{\alpha\beta\gamma\delta}(p,q,P)           \\
                  (231):  & \mathsf{F}'_n &= (\mathsf{F}_n)_{bcad}\,,  & \Psi'_n &= (\Psi_n)_{\beta\gamma\alpha\delta}(p',q',P)         \\
                  (312):  & \mathsf{F}''_n &= (\mathsf{F}_n)_{cabd}\,,  & \Psi''_n &= (\Psi_n)_{\gamma\alpha\beta\delta}(p'',q'',P)      \\[3mm]
                  (213):  & \widetilde{\mathsf{F}}_n &= (\mathsf{F}_n)_{bacd}\,,  & \widetilde{\Psi}_n &= (\Psi_n)_{\beta\alpha\gamma\delta}(p,-q,P)         \\
                  (321):  & \widetilde{\mathsf{F}}'_n &= (\mathsf{F}_n)_{cbad}\,,  & \widetilde{\Psi}'_n &= (\Psi_n)_{\gamma\beta\alpha\delta}(p',-q',P)      \\
                  (132):  & \widetilde{\mathsf{F}}''_n &= (\mathsf{F}_n)_{acbd}\,,  & \widetilde{\Psi}''_n &= (\Psi_n)_{\alpha\gamma\beta\delta}(p'',-q'',P)
             \end{array}
             \end{equation*}
             where for instance $(231)$ denotes a permutation $\{p_1, p_2, p_3\} \rightarrow \{ p_2, p_3, p_1\}$.
             The resulting permuted relative momenta read:
 \renewcommand{\arraystretch}{1.3}
             \begin{equation}\label{perm-momenta}
                 \begin{array}{rl}
                     p' &= -q-\frac{p}{2}\,, \\
                     q' &= -\frac{q}{2} + \frac{3p}{4}\,,
                 \end{array}\qquad
                 \begin{array}{rl}
                     p'' &= q-\frac{p}{2}\,, \\
                     q'' &= -\frac{q}{2} - \frac{3p}{4}\,.
                 \end{array}
             \end{equation}
             The Dirac and flavor doublets (for the latter: replace $\Psi$ by $\mathsf{F}$) transform under these permutations as
             \begin{equation}\label{perm-tf-doublet}
                 \Psi = \mathcal{M}' \Psi' = \mathcal{M}''\Psi'' = \widetilde{\mathcal{M}} \widetilde{\Psi} = \widetilde{\mathcal{M}}' \widetilde{\Psi}' = \widetilde{\mathcal{M}}'' \widetilde{\Psi}''\,,
             \end{equation}
             where the transformation matrices acting upon the index $n$ are given by $\widetilde{\mathcal{M}} = \text{diag}(-1,1)$ and
 \renewcommand{\arraystretch}{1.2}
             \begin{equation*}
             \begin{split}
                   \mathcal{M}' = \frac{1}{2}\left( \begin{array}{cc} -1 & -\sqrt{3} \\ \sqrt{3} & -1 \end{array}\right), \quad
                   \mathcal{M}'' = \frac{1}{2}\left( \begin{array}{cc} -1 & \sqrt{3} \\ -\sqrt{3} & -1 \end{array}\right),   \\
                   \widetilde{\mathcal{M}}' = \frac{1}{2}\left( \begin{array}{cc}  1 & -\sqrt{3} \\ -\sqrt{3} & -1 \end{array}\right), \quad
                   \widetilde{\mathcal{M}}'' = \frac{1}{2}\left( \begin{array}{cc} 1 & \sqrt{3} \\  \sqrt{3} & -1 \end{array}\right).   \\
             \end{split}
             \end{equation*}
             Because of their orthogonality a scalar product of two doublets,
             such as the bracket in Eq.~\eqref{FE:nucleon_amplitude_full}, transforms as a symmetric singlet
             under a permutation.

             It is convenient to define Lorentz-invariant momentum variables with definite transformation properties under the permutation group, cf.~Ref.~\cite{Carimalo:1992ia},
             a feature which is not manifest in the set $\{ p^2, q^2, z_0, z_1, z_2 \}$.
             Indeed one can construct a symmetric singlet
             \begin{equation}
                 x := \frac{p^2}{4} + \frac{q^2}{3} \in \mathds{R}_+\,,
             \end{equation}
             and two further doublets
             \begin{equation}\label{doublet-variables}
             \begin{split}
                 \left(\begin{array}{c} y_1 \\ y_2 \end{array} \right) &:= \frac{1}{1+\xi}\,\left(\begin{array}{c} \xi-1 \\ 2\,\sqrt{\xi}\,\hat{p}\cdot\hat{q} \end{array} \right)\\
                 \left(\begin{array}{c} w_1 \\ w_2 \end{array} \right) &:= \frac{1}{\sqrt{1+\xi}}\left(\begin{array}{c} z_1 \\ \sqrt{\xi}\,z_2 \end{array} \right)
             \end{split}
             \end{equation}
             on the domain $(-1,1)\in\mathds{R}$, with $\xi := (4q^2)/(3p^2)$ and
             \begin{equation}
                \hat{p}\cdot\hat{q} = z_1  z_2 + z_0 \sqrt{1-z_1^2} \sqrt{1-z_2^2}\,.
             \end{equation}
             Under a permutation of momenta \eqref{perm-momenta} the doublets again transform via Eq.~\eqref{perm-tf-doublet}.

        \subsection{Orthonormal basis} \label{sec:orthonormalbasis}

 \renewcommand{\arraystretch}{1.3}

            The mixed-(anti-)symmetric Dirac amplitudes $\Psi_n$ in Eq.~\eqref{FE:nucleon_amplitude_full} have the general structure of four-fermion Green functions
            with positive parity and positive energy. As such they are described by 64 covariant basis elements
            which we denote by
            \begin{equation} \label{Faddeev:orth-basis}
            \begin{split}
               \mathsf{X}_{1,ij\omega} &:= \mathsf{T}_{ij}  \left( \Lambda_\omega \gamma_5 C \otimes \Lambda_+ \right), \\
               \mathsf{X}_{2,ij\omega} &:= \left( \gamma_5 \otimes \gamma_5 \right) \mathsf{X}_{1,ij\omega}
            \end{split}
            \end{equation}
            where $\Lambda_\omega$ ($\omega=\pm$) are the positive- and negative-energy projectors from Eq.~\eqref{energy-projectors}, and
            restriction to positive energies for the nucleon eliminates the occurrence of $\Lambda_-$ on the right-hand side of the tensor product.
            The dependence on the relative momenta $p$ and $q$ is carried by the
            16 tensors $\mathsf{T}_{ij}$ with $i,j=1\dots 4$ which are detailed below. The
            Dirac amplitudes are then reconstructed via
            \begin{equation}\label{amplitude:reconstruction}
                \Psi_n(p,q,P) = \sum_{kij\omega} f_{n,kij\omega}(t)\,\mathsf{X}_{k,ij\omega}(r,s,\hat{P})\,,
            \end{equation}
            where the coefficients $f_{n,kij\omega}(t)$ depend on the 5 Lorentz-invariant
            momentum variables $t:=\{p^2,q^2,z_0,z_1,z_2\}$.
            It is convenient to express the momenta
            $\{p,q,P\}$ through orthogonal unit vectors $\{r,s,\hat{P}\}$ which satisfy
            $r^2 = s^2 = \hat{P}^2 = 1$ and $r\cdot s = r\cdot \hat{P} = s\cdot\hat{P} = 0$.
            This is achieved in a covariant way via
            \begin{equation*}
                r = \widehat{p_T} = \frac{\hat{p}-z_1 \hat{P}}{\sqrt{1-z_1^2}}\,, \quad
                \widehat{q_T} = \frac{\hat{q}-z_2 \hat{P}}{\sqrt{1-z_2^2}}\,, \quad
                s = \frac{\widehat{q_T}-z_0 \,r}{\sqrt{1-z_0^2}}\,
            \end{equation*}
            which, in the nucleon's rest frame ($\hat{P}=e_4$) and the coordinate representation~\eqref{APP:momentum-coordinates},
            (for $p^\mu$: replace $z\rightarrow z_1$ and $y\rightarrow 1$; for $q^\mu$: $p^2\rightarrow q^2$, $z\rightarrow z_2$, $y\rightarrow z_0$, $\phi\rightarrow 0$)
            assume the simple momentum-independent
            form $r=e_3$ and $s=e_2$.

            Let us now define the basic spin structures
            \begin{equation}\label{basic-spin-structures}
                \Gamma_{j=1\dots 4} = \{ \mathds{1}, \, \gamma_5 \,\Slash{v},\, \Slash{r},\,\Slash{s} \}\,,
            \end{equation}
            with $v^\mu = \epsilon^{\mu\alpha\beta}_T r^\alpha s^\beta$ and by virtue of Eq.~\eqref{epsilon-id-2}:
            $\gamma_5\,\Slash{v} = \Slash{r} \,\Slash{s} \,\hat{\Slash{P}}$.
            The assignment $\mathsf{T}_{ij} = \Gamma_i \otimes \Gamma_j$ would yield a
            complete but non-orthogonal basis~\eqref{Faddeev:orth-basis}.
            An orthonormal basis which also corresponds to a partial-wave decomposition in the nucleon's rest frame was presented in Ref.~\cite{Eichmann:2009en}. It reads
 \renewcommand{\arraystretch}{1.4}
             \begin{equation} \label{basis-spin-1/2}
                  \mathsf{T}_{1j} = \mathds{1}\otimes \Gamma_j\,, \qquad
                  \mathsf{T}_{2j} = \frac{1}{\sqrt{3}}\,\gamma_T^\alpha \otimes \gamma_T^\alpha \,\Gamma_j\,,
             \end{equation}
             with $\Gamma_j$ from Eq.~\eqref{basic-spin-structures}, and
             \begin{equation} \label{basis-spin-3/2}
                  \mathsf{T}_{3j} = \frac{1}{\sqrt{6}}\,\gamma_T^\alpha \otimes \Gamma^{\alpha}_j\,, \quad
                  \mathsf{T}_{4j} = \frac{1}{\sqrt{2}}\,\gamma_T^\alpha \otimes \gamma_T^\beta\,\Gamma^{\alpha\beta}_j\,,
             \end{equation}
             where $\Gamma^{\alpha}_j$, $\Gamma^{\alpha\beta}_j$ ($j=1\dots 4$) are defined by
             \begin{equation*}
                  \left[ \begin{array}{l}
                                         3 \,r^\alpha \Slash{r} - \gamma_T^\alpha  \\
                                         ( 3\,v^\alpha - \gamma_T^\alpha\,\Slash{v} ) \,\gamma_5 \\
                                         3\,r^\alpha - \gamma_T^\alpha\,\Slash{r} \\
                                         3\,s^\alpha - \gamma_T^\alpha\,\Slash{s}
                                             \end{array}\right] \;\text{and}\;
                 \left[ \begin{array}{l}
                                              r^\alpha r^\beta +2\,s^\alpha s^\beta - T_P^{\alpha\beta}  \\
                                              r^\alpha s^\beta + s^\alpha r^\beta  \\
                                               c_1^{\alpha\beta}\,\gamma_5  \\
                                               c_2^{\alpha\beta}\,\gamma_5
                                             \end{array} \right],
             \end{equation*}
             respectively, with
             \begin{equation}\label{nucleonbasis-c12}
             \begin{split}
                 c_1^{\alpha\beta} &= \left(2\, s^\alpha s^\mu - T_P^{\alpha\mu}\right) \epsilon^{\mu\beta\sigma}_T r^\sigma \,, \\
                 c_2^{\alpha\beta} &= \left(2\, r^\alpha r^\mu - T_P^{\alpha\mu}\right) \epsilon^{\mu\beta\sigma}_T s^\sigma\,.
             \end{split}
             \end{equation}
             The $\Gamma^{\alpha}_j$ and $\Gamma^{\alpha\beta}_j$ satisfy the relation
             \begin{equation}\label{spin-3/2-relation}
                 \gamma_T^\alpha\,\Gamma_j^\alpha = \gamma_T^\alpha\,\gamma_T^\beta\,\Gamma^{\alpha\beta}_j = 0\,.
             \end{equation}

             With this choice, the basis elements in Eq.~\eqref{Faddeev:orth-basis} obey the orthogonality relation
             \begin{equation}\label{basis:orthogonality}
                 \frac{1}{4}\,\text{Tr}\,\left\{ \conjg{\mathsf{X}}_{k,ij\omega}\,\mathsf{X}_{k',i'j'\omega'} \right\} = \delta_{kk'} \delta_{ii'} \delta_{jj'} \delta_{\omega\omega'}\,.
             \end{equation}
             Moreover, they are eigenstates of the total quark spin $\vect{S}^2$ and orbital angular momentum $\vect{L}^2$ with eigenvalues $s(s+1)$ and $l(l+1)$.
             The $s=\nicefrac{1}{2}$ subspace is spanned by the elements in Eq.~\eqref{basis-spin-1/2}; the $s=\nicefrac{3}{2}$ subspace by those in Eq.~\eqref{basis-spin-3/2}.
             $\mathsf{T}_{11}$ and $\mathsf{T}_{21}$ are $s-$waves ($l=0$), $\mathsf{T}_{31}$ and $\mathsf{T}_{4j}$ with $j=1\dots 4$ are $d-$waves ($l=2$)
             and the remaining ones are $p-$waves ($l=1$). We will carry out the partial-wave decomposition in detail in the next subsection.

             We note that the $\mathsf{T}_{ij}$ of Eqs.~(\ref{basis-spin-1/2}--\ref{basis-spin-3/2}) also constitute a complete orthonormal basis of a four-quark Green function with positive parity,
             however with Eq.~\eqref{Faddeev:orth-basis} generalized to the form
            \begin{equation}
            \begin{split}
               \mathsf{X}_{1,ij\omega\omega'} &:= \mathsf{T}_{ij}  \left( \Lambda_\omega \otimes \Lambda_{\omega'} \right), \\
                \mathsf{X}_{2,ij\omega\omega'} &:= \left( \gamma_5 \otimes \gamma_5 \right) \mathsf{X}_{1,ij\omega\omega'}
            \end{split}
            \end{equation}
            where we dropped $\gamma_5 C$ to indicate the case of a quark-antiquark four-point function.

        \subsection{Partial-wave decomposition} \label{sec:partialwave}

             The Dirac basis elements $\mathsf{X}_{k,ij\omega}$ of Eq.~\eqref{Faddeev:orth-basis}
             can be classified with respect to their quark-spin and orbital angular momentum content in the nucleon's rest frame
             which we will demonstrate explicitly in this appendix.
             Only the total angular momentum $j=1/2$ of the nucleon is Poincar\'e-invariant  while the interpretation in terms of total quark spin
             and orbital angular momentum will differ in every frame.
             The spin is described by the Pauli-Lubanski operator:
             \begin{equation}
                 W^\mu = \frac{1}{2} \,\epsilon^{\mu\alpha\beta}_T  J^{\alpha\beta}\,,
             \end{equation}
             where we chose the total momentum $P$ to be normalized, cf.~Eq.~\eqref{transverse-epsilon}.
             As generators of the Poincar\'e algebra, $J^{\mu\nu}$ and $P^\mu$ satisfy the usual commutation relations.
             The eigenvalues of the square of the Pauli-Lubanski operator,
             \begin{equation}
                 W^2 = \frac{1}{2} \,T_P^{\mu\alpha}\,T_P^{\nu\beta} J^{\mu\nu}J^{\alpha\beta} \longrightarrow j(j+1)
             \end{equation}
             define the spin $j$ of the particle.
             For a system of three particles with total momentum $P$ and relative momenta $p$ and $q$,
             the total angular momentum operator consists of the total quark spin $\vect{S}$ and the relative orbital angular momentum $\vect{L}=\vect{L}_{(p)}+\vect{L}_{(q)}$.
             Upon subsuming them into Lorentz-covariant operators
             \begin{align}
                 S^{\mu} &= \textstyle\frac{1}{4} \displaystyle \epsilon^{\mu\alpha\beta}_T
                            \left( \sigma^{\alpha\beta} \otimes \mathds{1} \otimes \mathds{1} +  \text{perm.} \right), \nonumber \\
                 L_{(p)}^{\mu} &= \textstyle\frac{i}{2} \displaystyle \epsilon^{\mu\alpha\beta}_T \left( p^\alpha \partial_p^\beta - p^\beta \partial_p^\alpha \right) \mathds{1} \otimes \mathds{1} \otimes \mathds{1}, \\
                 L_{(q)}^{\mu} &= \textstyle\frac{i}{2} \displaystyle \epsilon^{\mu\alpha\beta}_T \left( q^\alpha \partial_q^\beta - q^\beta \partial_q^\alpha \right) \mathds{1} \otimes \mathds{1} \otimes \mathds{1}  \nonumber
             \end{align}
             with $W^\mu = S^\mu + L_{(p)}^\mu + L_{(q)}^\mu$,
            one can show that the orthonormal basis elements of Eq.~\eqref{Faddeev:orth-basis} are
            eigenstates of the operators
            \begin{equation}\label{operators-1}
              S^2  = \textstyle\frac{9}{4} \displaystyle \,\mathds{1} \otimes \mathds{1} \otimes \mathds{1} + \textstyle\frac{1}{4} \displaystyle \left( \sigma_T^{\mu\nu} \otimes \sigma_T^{\mu\nu} \otimes \mathds{1} + \text{perm.} \right)\,,
            \end{equation}
            and $L^2 = L_{(p)}^2 + L_{(q)}^2 + 2\,L_{(p)}\!\cdot\! L_{(q)}$, where
             \begin{align}
                 L_{(p)}^2 &= 2  \, p_T\cdot\partial_p + \left(p_T^\mu \,p_T^\nu - p_T^2 \,T^{\mu\nu}_P \right) \partial_p^\mu \,\partial_p^\nu \,,\nonumber\\[0.1cm]
                 L_{(q)}^2 &= 2  \, q_T\cdot\partial_q + \left(q_T^\mu \,q_T^\nu - q_T^2 \,T^{\mu\nu}_P \right) \partial_q^\mu \,\partial_q^\nu \,, \label{operators-2}\\[0.1cm]
                 L_{(p)}\!\cdot\! L_{(q)} &=  \left( p_T^\mu\,q_T^\nu - p_T  \cdot  q_T\,\,T^{\mu\nu}_P \right) \partial_p^\nu \,\partial_q^\mu\,,\nonumber
             \end{align}
            with eigenvalues $s(s+1)$ and $l(l+1)$.

            In the nucleon's rest frame, where $\hat{P}=e_4$, the operators (\ref{operators-1}--\ref{operators-2}) assume the meaning of
             total quark spin and orbital angular momentum.
            Here $S^2$ reads explicitly
            \begin{equation}
               S^2 =  \textstyle\frac{9}{4} \displaystyle\big( \mathds{1} \otimes \mathds{1} \otimes \mathds{1} \big)
                                    + \textstyle\frac{1}{2} \big( \mathbf{\Sigma} \otimes \mathbf{\Sigma} \otimes \mathds{1} +\text{perm.} \big)
            \end{equation}
            where $\Sigma_i = \frac{1}{2}\,\varepsilon_{ijk}\,\sigma_{jk}$. Choosing hyperspherical coordinates $\{p^2, \,z, \,y, \,\phi\}$ for $p^\mu$
            and $\{q^2, \,z', \,y', \,\phi'\}$ for $q^\mu$ according to Eq.~\eqref{APP:momentum-coordinates},
 \renewcommand{\arraystretch}{1.0}
            defining $\vect{p}$, $\vect{q}$ as the three-vectors corresponding to $p_T$ and $q_T$, e.g.
            \begin{equation}\label{APP:momentum-coordinates-Restframe}
                \vect{p} = \sqrt{p^2}\,\sqrt{1-z^2} \left( \begin{array}{c} \sqrt{1-y^2}\,\sin{\phi} \\
                                                                            \sqrt{1-y^2}\,\cos{\phi} \\
                                                                            y   \end{array}\right),
            \end{equation}
            and expressing the gradient and Laplacian in spherical coordinates, 
            the operator $L^2$ is the combination of
 \renewcommand{\arraystretch}{1.4}
             \begin{align*}
                 L_{(p)}^2 =&\;  2 \,\vect{p} \cdot\grad_{\vect{p}}  +p^k (\vect{p} \cdot\grad_{\vect{p}}) \,\nabla^k_{\vect{p}}  -\mathbf{p}^2 \Delta_{\vect{p}} \\
                                     =&\;  2y \,\partial_{y} -(1-y^2)\,\partial_{y}^2 -\frac{\partial_{\phi}^2}{1-y^2}\,, \\
                 L_{(q)}^2 =&\;  2 \,\vect{q} \cdot\grad_{\vect{q}}  +q^k (\vect{q} \cdot\grad_{\vect{q}}) \,\nabla^k_{\vect{q}}  -\mathbf{q}^2 \Delta_{\vect{q}} \\
                                     =&\;  2y' \,\partial_{y'} -(1-{y'}^2)\,\partial_{y'}^2 -\frac{\partial_{\phi'}^2}{1-{y'}^2}\,,
             \end{align*}
             and
             \begin{align*}
                 L_{(p)}\!\cdot\! L_{(q)} =&\; p^k (\vect{q} \cdot\grad_{\vect{p}}) \,\nabla^k_{\vect{q}} -(\vect{p}\cdot\vect{q}) (\grad_{\vect{p}}\cdot\grad_{\vect{q}}) = \\
                                          =&\;  -\cos(\phi-\phi') \sqrt{1-y^2}\,\sqrt{1-{y'}^2}\,\partial_y\,\partial_{y'} \\
                                           &\; -\left[ 1 + \frac{y y'\,\cos(\phi-\phi')}{\sqrt{1-y^2}\,\sqrt{1-{y'}^2}}  \right] \partial_\phi\,\partial_{\phi'}  \\
                                           &\; +\frac{\sin(\phi-\phi')}{\sqrt{1-y^2}\,\sqrt{1-{y'}^2}} \times \\
                                           &\; \times \left[y'\,(1-y^2)\,\partial_y\,\partial_{\phi'}  - y\, (1-{y'}^2)\,\partial_{y'}\,\partial_\phi\right].
             \end{align*}

            While these expressions are expedient, e.g., for use with a symbolic programming language, Eqs.~(\ref{operators-1}--\ref{operators-2})
            allow for a partial-wave decomposition in a more direct way.
            To find the eigenstates of $S^2$ one substitutes in Eq.~\eqref{operators-1} the relation
            \begin{equation}
                \sigma^{\mu\nu}_T \otimes \sigma^{\mu\nu}_T = -2\,\gamma^\mu_T\,\gamma_5 \hat{\Slash{P}} \otimes \gamma^\mu_T\,\gamma_5 \hat{\Slash{P}}
            \end{equation}
            that follows from Eq.\,\eqref{epsilon-id-2}. Then, with the anticommutation properties of $\gamma_T^\alpha$, $\gamma_5$ and $\hat{\Slash{P}}$
            and repeated use of Eq.\,\eqref{epsilon-id-2}
            it is straightforward to show that
            \begin{equation*}
                \left(S^2\right)_{\alpha\alpha'\beta\beta'\gamma\gamma'} \left( \mathsf{X}_{k,ij\omega} \right)_{\alpha'\beta'\gamma'\delta} = s(s+1)\,\left( \mathsf{X}_{k,ij\omega} \right)_{\alpha\beta\gamma\delta}\,,
            \end{equation*}
            with $s=\nicefrac{1}{2}$ for $i=1,2$ and $s=\nicefrac{3}{2}$ for $i=3,4$.
            The proof only relies on the generic form~(\ref{basis-spin-1/2}--\ref{basis-spin-3/2}) of the $\mathsf{T}_{ij}$, together with the property~\eqref{spin-3/2-relation},
            but not the specific form of the matrices $\Gamma_j$, $\Gamma^\alpha_j$ and $\Gamma^{\alpha\beta}_j$.

           The eigenstates of $L^2$ can be determined independently of the Dirac structure since
           the operators
           $L_{(p)}^2$, $L_{(q)}^2$ and $L_{(p)}\!\cdot\! L_{(q)}$ only act upon the relative-momentum dependence.
           The elements $\mathsf{T}_{11}$ and $\mathsf{T}_{21}$ are independent of the relative momenta, hence
           they carry orbital angular momentum $l=0$.
           All other basis elements in (\ref{basis-spin-1/2}--\ref{basis-spin-3/2})
           can be expressed through contractions of
           \begin{equation}\label{rs-1}
           \begin{split}
              r^\alpha, \quad
              s^\alpha, \quad &
              r^\alpha \, s^\beta, \quad
              s^\alpha \, r^\beta, \quad
              r^\alpha \, r^\beta, \quad
              s^\alpha \, s^\beta, \\
              &
              r^\alpha \, r^\beta \, s^\gamma, \quad
              s^\alpha \, s^\beta \, r^\gamma
           \end{split}
           \end{equation}
           with appropriate Dirac structures, for instance: $\mathsf{T}_{13} = (\mathds{1}\otimes \gamma_T^\alpha)\,r^\alpha$.
           Applying the operator $L^2$ on \eqref{rs-1} yields for example
           \begin{equation}\label{apply-L2}
           \begin{split}
               L^2 \,r^\alpha &= 2\,r^\alpha, \\
               L^2 \,s^\alpha &= 2\,s^\alpha, \\
               L^2\, r^\alpha r^\beta &= 6\,r^\alpha r^\beta - 2\,T_P^{\alpha\beta},\\
               L^2\, s^\alpha s^\beta &= 6\,s^\alpha s^\beta - 2\,T_P^{\alpha\beta},\\
               L^2\, r^\alpha s^\beta &= 4\,r^\alpha s^\beta  +2\,s^\alpha r^\beta\,,
           \end{split}
           \end{equation}
           and the results for higher powers of momenta follow from the relation
           \begin{equation}
           \begin{split}
              L^2\,(fgh) &= f\,L^2\,(gh) + g\,L^2\,(fh) + h\,L^2\,(fg)  \\ & - (gh)\,L^2\,f - (fh)\,L^2\,g - (fg)\,L^2\,h\,.
           \end{split}
           \end{equation}
           From~\eqref{apply-L2} one constructs the eigenfunctions for $l=1$:
           \begin{equation}
               r^\alpha, \quad
               s^\alpha, \quad
               r^\alpha s^\beta - s^\alpha r^\beta = v^\mu\,\epsilon^{\mu\alpha\beta}_T \,.
           \end{equation}
           The unit vectors $r^\alpha$, $s^\alpha$ and $v^\alpha$ appear in the basis elements $\mathsf{T}_{ij}$ for $i=1,2,3$ and $j=2,3,4$ which
           therefore carry $l=1$.
           For $l=2$ one obtains the eigenfunctions
           \begin{equation}
                 r^\alpha s^\beta + s^\alpha r^\beta, \quad
                 3\,r^\alpha r^\beta - T_P^{\alpha\beta}, \quad
                 3\,s^\alpha s^\beta - T_P^{\alpha\beta}
           \end{equation}
           that appear in $\mathsf{T}_{31}$, $\mathsf{T}_{41}$ and $\mathsf{T}_{42}$, and also the eigenfunctions
           \begin{align}
               & \mathcal{A}_{[\beta\gamma]} \,( 2\,s^\alpha s^\beta - T_P^{\alpha\beta} )\,r^\gamma = -c_1^{\alpha\mu}\, \epsilon^{\mu\beta\gamma}_T\,, \\
               & \mathcal{A}_{[\beta\gamma]} \,( 2\,r^\alpha r^\beta - T_P^{\alpha\beta} )\,s^\gamma = -c_2^{\alpha\mu}\, \epsilon^{\mu\beta\gamma}_T
           \end{align}
           which occur in $\mathsf{T}_{43}$ and $\mathsf{T}_{44}$.
           Here $\mathcal{A}_{[\beta\gamma]}$ denotes an antisymmetrizer with respect to the indices $\beta$, $\gamma$
           and the $c_{1,2}^{\alpha\beta}$ are defined in Eq.~\eqref{nucleonbasis-c12}.
           A second set of eigenfunctions with $\mathcal{A}_{[\alpha\gamma]}$ instead of $\mathcal{A}_{[\beta\gamma]}$ yields no new information.

           To summarize, the 16 elements $\mathsf{T}_{ij}$ can be categorized into two $s-$waves, nine $p-$waves and five $d-$waves.
           In the final basis elements of Eq.~\eqref{Faddeev:orth-basis} these numbers are multiplied by four through the additional indices $k=1,2$ and $\omega=\pm$.

  \section{Faddeev equation: updated solution strategy} \label{sec:faddeeveqsolution}

            In this appendix we detail an updated solution strategy for the Faddeev equation which considerably reduces the numerical effort.
            We start by writing the Faddeev equation for the Dirac part of the nucleon amplitude as
            \begin{equation}
               \Psi_n = \Psi_n^{(1)} + \Psi_n^{(2)} + \Psi_n^{(3)}
            \end{equation}
            where $\Psi_n = (\Psi_n)_{\alpha\beta\gamma\delta}(p,q,P)$ is the Dirac doublet in Eq.~\eqref{FE:nucleon_amplitude_full}
            involving the mixed-antisymmetric ($\Psi_1$) and mixed-symmetric ($\Psi_2$) contributions,
            and the $\Psi_n^{(a)}$ represent the
            three diagrams in the Faddeev equation~\eqref{faddeev:eq}.
            We write the amplitude decomposition as
            \begin{equation}\label{amplitude-reconstruction-2}
                \Psi_n(p,q,P) = \sum_i f_{n,i}(t)\,\mathsf{X}_i(r,s,\hat{P})\,,
            \end{equation}
            where the index $i$ now collects all the previous indices $\{k,i,j,\omega\}$ from Eq.~\eqref{amplitude:reconstruction},
            and $t=\{p^2,q^2,z_0,z_1,z_2\}$ again abbreviates the invariant momentum variables.
            Using the trace orthogonality \eqref{basis:orthogonality} yields
            equations for the amplitude dressing functions:
            \begin{equation}\label{FE-for-dressingfunctions-1}
                f_{n,i}(t) = \sum_{a=1}^3 f_{n,i}^{(a)}(t) \,,
            \end{equation}
            where the contributions on the r.h.s. are given by
            \begin{equation}\label{FE-for-dressingfunctions-2}
            \begin{split}
                f_{n,i}^{(a)}(t) &= \int\limits_k \mathcal{K}_{ij}^{(a)} (p,q,k,P)\,g_{n,j}^{(a)}(t^{(a)})\,, \\
                g_{n,i}^{(a)}(t) &=  \mathcal{G}_{ij}^{(a)} (t)\,f_{n,j}(t)\,.
            \end{split}
            \end{equation}
            Here the $g_{n,i}^{(a)}$ denote the coefficients of the wave functions $\Phi_n^{(a)}=S(p_b) S(p_c) \Psi_n$ with attached propagator legs,
            where $\{a,b,c\}$ is a symmetric permutation of $\{1,2,3\}$, with the same Poincar\'e-covariant decomposition as in Eq.~\eqref{amplitude-reconstruction-2}.
            The internal momentum variables $t^{(a)}$ are obtained from the internal momenta of Eq.~\eqref{fe-momenta-internal}.
            The kernel and propagator matrices in \eqref{FE-for-dressingfunctions-2} follow from a projection
            onto the basis elements:
            \begin{equation}
            \begin{split}
                \mathcal{K}_{ij}^{(3)} = \;& \textstyle{\frac{1}{4}}\,\conjg{\mathsf{X}}_i(r,s,\hat{P})_{\beta\alpha\delta\gamma}\, \mathsf{X}_j(r^{(3)},s^{(3)},\hat{P})_{\alpha'\beta'\gamma\delta}   \\
                &  \times K_{\alpha\alpha'\beta\beta'}(k)\,, \\
                \mathcal{G}_{ij}^{(3)} = \;& \textstyle{\frac{1}{4}}\,\conjg{\mathsf{X}}_i(r,s,\hat{P})_{\beta\alpha\delta\gamma}\, \mathsf{X}_j(r,s,\hat{P})_{\alpha'\beta'\gamma\delta}   \\
                &  \times S_{\alpha\alpha'}(p_1)\,S_{\beta\beta'}(p_2)\,, \\
            \end{split}
            \end{equation}
            and a cyclic permutation of the index pairs and quark momenta in
            the second and fourth row yields the remaining expressions for $a=1,2$.

            Eqs.~(\ref{FE-for-dressingfunctions-1}--\ref{FE-for-dressingfunctions-2}) can be solved by iteration which requires a compromise between huge memory requirements for the kernel
            and propagator matrices (if computed in advance) or large CPU times (if computed in each iteration step anew).
            Compared to the permuted diagrams for $a=1,2$, the kernel $\mathcal{K}^{(3)}$ is less difficult to handle:
            in the kinematics of Eq.~\eqref{fe-momenta-internal} the internal relative momentum $p^{(3)}$ equals
            the external one $p$, and hence $r^{(3)}=r$ which in the rest frame reduces to $e_3$.  
            In combination with the simplicity of the rainbow-ladder kernel which only depends on the gluon momentum $k=q^{(3)}-q$,
            the kernel $\mathcal{K}^{(3)}$ is independent of the variables $p^2$ and $z_1$ and hence 
            only requires a modest amount of memory (typically several GB).

            Nevertheless the storage problem for $\mathcal{K}^{(1)}$ and $\mathcal{K}^{(2)}$ remains.
            In this respect it is beneficial to take advantage of the permutation-group properties of the amplitude.
            Applying the relations of Eq.~\eqref{perm-tf-doublet} for the internal amplitudes $\Psi(p^{(1)},q^{(1)},P)$ and $\Psi(p^{(2)},q^{(2)},P)$ in the Faddeev equation~\eqref{faddeev:eq},
            and expressing the relative momenta in each diagram by $p'$, $q'$ or $p''$, $q''$, respectively,
            yields the relation
            \begin{equation} \label{FE:newmethod}
               \Psi = \Psi^{(3)} + \mathcal{M}'\big[\Psi^{(3)}\big]' + \mathcal{M}''\big[\Psi^{(3)}\big]''\,.
            \end{equation}
            By means of Eq.~\eqref{FE:newmethod}, the amplitudes $\Psi^{(1)}$ and $\Psi^{(2)}$ no longer need to be computed explicitly but can
            be reconstructed from $\Psi^{(3)}$.
            Note that the full Dirac-color-flavor amplitude corresponding to $\Psi^{(3)}$ is not totally antisymmetric,
            i.e. its Dirac parts do not transform via Eq.~\eqref{perm-tf-doublet}, and the right-hand side
            of the above equation is not simply $3\,\Psi^{(3)}$.

            The orthogonality relation \eqref{basis:orthogonality} then yields
            corresponding equations for the amplitude dressing functions:
            \begin{equation}\label{FE:newmethod-for-dressingfunctions}
                f_i(t) = f_i^{(3)}(t) + \mathcal{M}'H_{ij}'\,f_j^{(3)}(t') + \mathcal{M}''H_{ij}''\,f_j^{(3)}(t'')\,
            \end{equation}
            where $t'$ and $t''$ are the Lorentz-invariant momentum variables
            obtained from the permuted momenta \eqref{perm-momenta};
            $\mathcal{M}'$ and $\mathcal{M}''$ act on the doublet index $n$; and the matrices $H'$ and $H''$ are given by
            \begin{equation}
            \begin{split}
                H'_{ij}(t) &= \textstyle{\frac{1}{4}}\,\text{Tr}\,\left\{ \conjg{\mathsf{X}}_i(r,s,\hat{P})\,\mathsf{X}'_j(r',s',\hat{P}) \right\}, \\
                H''_{ij}(t) &= \textstyle{\frac{1}{4}}\,\text{Tr}\,\left\{ \conjg{\mathsf{X}}_i(r,s,\hat{P})\,\mathsf{X}''_j(r'',s'',\hat{P}) \right\},
            \end{split}
            \end{equation}
            where $\mathsf{X}'_j$, $\mathsf{X}''_j$ are the basis elements with permuted Dirac indices according to App.~\ref{sec:permutationgroup},
            and the unit vectors $r'$, $r''$, $s'$, $s''$ follow from the permuted momenta in Eq.~\eqref{perm-momenta}.
            Upon exchanging $p^2$ and $q^2$ with the singlet and doublet variables $x$ and $y_1$ of Eq.~\eqref{doublet-variables} one finds that
            $H'$ and $H''$ are independent of $x$; and by taking into account the sparseness of $\mathcal{K}_{ij}^{(3)}$
            and $H'$, $H''$, these matrices can be stored in advance.
            Instead of interpolating the dressing functions for the permuted momentum variables in each iteration step again,
            one may expand them in orthogonal polynomials.
            Due to the weak angular dependencies in $z_0$, $z_1$ and $z_2$ (and also $y_1$) sufficient accuracy
            is already reached by retaining a small number of moments.

            The algorithm for solving the Faddeev equation can now be summarized as follows:
            start with a guess for $f_{n,i}(t)$; in each iteration step, compute $f_{n,i}^{(3)}(t)$ from Eq.~\eqref{FE-for-dressingfunctions-2} and $f_{n,i}(t)$ from Eq.~\eqref{FE:newmethod-for-dressingfunctions};
            proceed until converged. These steps are repeated for different nucleon test masses for which purpose an eigenvalue $\lambda(M)$ is introduced
            in the equation; the correct nucleon mass yields the eigenvalue $\lambda(M)=1$.
            With the procedure detailed here, solving the Faddeev equation on a cluster (or even a potent desktop computer),
            \textit{without} any approximation on the momentum dependence
            and with accuracy that is sufficient to determine a reliable nucleon mass,
            becomes possible within a few hours.

  \section{Nucleon electromagnetic current} \label{sec:current}

        \subsection{General properties} \label{sec:currentgeneral}

             The matrix-valued electromagnetic current of the nucleon can be written in the most general form as
             \begin{equation}\label{eq:current-general1}
                 J^{\mu}(P,Q) = \Lambda_+^f\left( (F_1+F_2)\,i\gamma^\mu - F_2\,\frac{P^\mu}{M}\right)\Lambda_+^i\,,
             \end{equation}
             where $\mu$ is the photon index and $J_{\delta'\delta}^\mu(P,Q)$ is a Dirac matrix with indices
             $\delta$, $\delta'$ for incoming and outgoing nucleon amplitudes.
             We abbreviated the positive-energy projectors by $\Lambda_+(P_{f,i}) = \Lambda_+^{f,i}$.
             The current involves two momenta, expressed through the incoming and outgoing momenta $P_i$, $P_f$ or by the
             average momentum $P=(P_i+P_f)/2$ and photon momentum $Q=P_f-P_i$. Since the nucleon is on-shell, $P_i^2=P_f^2=-M^2$, one has
             \begin{equation}
                 P^2 = -M^2 (1+\tau), \qquad P\cdot Q=0\,,
             \end{equation}
             where $\tau := Q^2/(4M^2)$; hence the Lorentz-invariant form factors which constitute the vertex
             can only depend on the photon momentum-transfer $Q^2$.
             Contracting Eq.~\eqref{eq:current-general1} with nucleon spinors $\conjg{u}(P_f,s_f)$, $u(P_i,s_i)$
             which are eigenstates of $\Lambda_+^{f,i}$, e.g.:
             \begin{equation}
                 \Lambda_+^i \,u(P_i,s) = u(P_i,s)\,, \quad
                 s = \pm\nicefrac{1}{2} \,,
             \end{equation}
             yields the current-matrix element
             $\langle P_f, s_f \,|\, J^\mu\,| \,P_i, s_i \rangle$.

             As a three-point function depending on two momenta, with one vector and two spinor legs,
             the bracket in Eq.\,\eqref{eq:current-general1} could in principle involve the 12 positive-parity tensor structures
             \begin{equation} \label{spinorstructures}
                 \left\{ \, \gamma^\mu,\, P^\mu,\, Q^\mu \, \right\} \times \left\{ \, \mathds{1},\, \Slash{P},\, \Slash{Q},\, \left[\Slash{P}, \Slash{Q}\right] \, \right\}.
             \end{equation}
             Applying the positive-energy projectors reduces this set via Eq.~\eqref{energy-projectors-properties} to the three basis elements $\left\{ \, \gamma^\mu,\, P^\mu,\, Q^\mu \, \right\}$,
             and imposing charge-conjugation invariance of the current, $ \conjg{J^{\mu}}(P,Q)  \stackrel{!}{=} J^{\mu}(P,-Q)$, eliminates the component $Q^\mu$.
             The resulting current is automatically conserved, i.e. $Q^\mu J^{\mu} = 0$.
             Consequentially, the general electromagnetic current of the nucleon depends only on two form factors $F_i(Q^2)$.
             Using the Gordon identity
             \begin{equation}\label{eq:gordonid}
                 \Lambda_+^f \left[ \gamma^\mu + \frac{iP^\mu}{M} + \frac{\sigma^{\mu\nu}Q^\nu}{2M}\right]\Lambda_+^i = 0
             \end{equation}
             finally leads to the expression given in Eq.~\eqref{eq:current2},
             and the electric and magnetic form factors are extracted via the Dirac traces in Eq.~\eqref{obtain-ffs}.

        \subsection{Diagrams in the three-quark framework} \label{sec:emcurrent-worked-out}

            In this appendix we collect the ingredients of the nucleon electromagnetic current operator
            in the three-quark framework which is depicted in Fig.~\ref{fig:faddeev-current-specific}.
            In a rainbow-ladder truncation
             the current is the sum of the impulse-approximation and kernel diagram, the sum of the three permutations $a=1,2,3$,
             and the sum of the mixed-antisymmetric and mixed-symmetric components in the incoming ($n=1,2$) and outgoing ($n'=1,2$) nucleon amplitudes:
             \begin{equation}\label{current:total}
                 J^\mu_{\delta'\delta}(P,Q) = \sum_{a=1}^3 \sum_{n'n} \left[ J^{(a),\text{IMP}}_{n'n} + J^{(a),\text{K}}_{n'n}\right]^\mu_{\delta'\delta}\,.
             \end{equation}
             For example, the impulse-approximation diagram where the photon couples to the upper quark leg reads explicitly:
             \begin{equation}\label{current:diagram-3}
             \begin{split}
                 &\left[ J^{(3),\text{IMP}}_{n'n} \right]^\mu_{\delta'\delta} = \mathsf{F}^{(3)}_{n'n}\, \int\limits_p \!\!\!\int\limits_q (\conjg{\Psi}_{n'})_{\beta'\alpha'\delta'\gamma'}(p_f,q_f,P_f) \,\times \\
                 & \times S_{\alpha'\alpha}(p_1)\,S_{\beta'\beta}(p_2)\,\left[ S(p_3^+)\,\Gamma^\mu_\text{q}(p_3,Q)\,S(p_3^-)\right]_{\gamma'\gamma} \times \\
                 & \times (\Psi_n)_{\alpha\beta\gamma\delta}(p_i,q_i,P_i)  \,,
             \end{split}
             \end{equation}
             where $p_i$, $q_i$ and $p_f$, $q_f$ are the incoming and outgoing relative momenta;
             $P_i$ and $P_f$ are the incoming and outgoing nucleon momenta;
             $p_1$, $p_2$, $p_3$ and $p_3^\pm=p_3 \pm Q/2$ are the quark momenta;
             $S(p_i)$ are the dressed-quark propagators;
             $\Gamma^\mu_\text{q}$ is the dressed quark-photon vertex; and $p$ and $q$ are the two loop momenta.
             For symmetric momentum partitioning (i.e., a momentum-partitioning parameter $\nicefrac{1}{3}$) the relative momenta are explicitly given by
             \begin{equation}\label{ff-momenta-1}
                 p_f = p + \frac{Q}{3}\,, \quad
                 p_i = p - \frac{Q}{3}\,, \quad
                 q_f = q_i = q
             \end{equation}
             and the quark momenta by
             \begin{equation}\label{ff-momenta-2}
                 p_1 = -q - \frac{p}{2} + \frac{P}{3}\,, \quad
                 p_2 =  q - \frac{p}{2} + \frac{P}{3}
             \end{equation}
             and $p_3 = p+P/3$.

             The flavor trace in Eq.~\eqref{current:diagram-3} denotes
             \begin{equation} \label{flavor-current}
                 \mathsf{F}^{(3)}_{n'n} = (\mathsf{e_i})_{d'} \, (\mathsf{F}^\dag_{n'})_{bad'c'} \,\mathsf{Q}_{c'c} \,(\mathsf{F}_n)_{abcd}\,(\mathsf{e_i})_d\,,
             \end{equation}
             where $\mathsf{Q} = \text{diag} (q_u,q_d)$ is the quark charge matrix attached to the quark-photon vertex
             and $\mathsf{F}_n$, $\mathsf{F}^\dag_{n'}$ are the flavor matrices of Eq.~\eqref{FAD:flavor}.
             Applying the isospin vectors $\mathsf{e_1}=(1,0)$ or $\mathsf{e_2}=(0,1)$ in Eq.~\eqref{flavor-current} singles out the proton and neutron contributions, respectively.
             This yields
    \renewcommand{\arraystretch}{1.0}
             \begin{equation}
                 \mathsf{F}^{(3)}  = \left( \begin{array}{cc}
                                            q_u & 0 \\
                                            0 & \frac{1}{3} \,(q_u+2 q_d)
                                            \end{array}  \right) =
                                          \frac{2}{3}\left(
                                            \begin{array}{cc}
                                              1 & 0 \\
                                              0 & 0 \\
                                            \end{array}
                                          \right)
             \end{equation}
             for the proton and
             \begin{equation}
                 \mathsf{F}^{(3)}  = \left( \begin{array}{cc}
                                            q_d & 0 \\
                                            0 & \frac{1}{3} \,(q_d+2 q_u)
                                            \end{array}  \right) =
                                          -\frac{1}{3}\left(
                                            \begin{array}{cc}
                                              1 & 0 \\
                                              0 & -1 \\
                                            \end{array}
                                          \right)
             \end{equation}
             for the neutron flavor traces.
             The color traces for the impulse-approximation diagrams equal $-1$.

             The current diagram involving the rainbow-ladder kernel has the same overall shape as Eq.~\eqref{current:diagram-3},
             \begin{equation}\label{current:diagram-kernel-3}
             \begin{split}
                 &\left[ J^{(3),\text{K}}_{n'n} \right]^\mu_{\delta'\delta} = \mathsf{F}^{(3)}_{n'n}\,\int\limits_p \!\!\!\int\limits_q (\conjg{\Psi}_{n'})_{\beta'\alpha'\delta'\gamma'}(p_f,q_f,P_f) \,\times \\
                 & \times S_{\alpha'\alpha}(p_1)\,S_{\beta'\beta}(p_2)\,\left[ S(p_3^+)\,\Gamma^\mu_\text{q}(p_3,Q)\,S(p_3^-)\right]_{\gamma'\gamma} \times \\
                 & \times (\Psi_n^{(3)})_{\alpha\beta\gamma\delta}(p_i,q_i,P_i)  \,,
             \end{split}
             \end{equation}
             except for a color factor $\nicefrac{2}{3}$ and an incoming amplitude which is replaced by:
             \begin{equation*}
             \begin{split}
                 &(\Psi_n^{(3)})_{\alpha\beta\gamma\delta}(p_i,q_i,P_i) =  \int\limits_k K_{\alpha\alpha'\beta\beta'}(k)\,\times \\
                 & \times S_{\alpha'\alpha''}(p_1-k)\, S_{\beta'\beta''}(p_2+k) \, (\Psi_n)_{\alpha''\beta''\gamma\delta}(p_i,q_i+k,P_i) \,.
             \end{split}
             \end{equation*}
             The practical form-factor calculation can be simplified by writing the sum of Eqs.~\eqref{current:diagram-3} and \eqref{current:diagram-kernel-3}
             schematically as
             \begin{equation*}
             \begin{split}
                 J^{(3),\text{SUM}} &= \int\!\!\!\int \conjg{\Psi}_f\,S_1 S_2 \,(S_3 \,\Gamma^\mu S_3) \,(\Psi_i -\Psi_i^{(3)})  \\
                         &= \int\!\!\!\int \conjg{\Psi}_f\,(S_3 \,\Gamma^\mu)\,(S_1 S_2 S_3)\,( \Psi_i -\Psi_i^{(3)})   \\
                         &=\int\limits_p (S_3\,\Gamma^\mu) \int\limits_q \conjg{\Psi}_f\,(\Phi_i-\Phi_i^{(3)}) \,,
             \end{split}
             \end{equation*}
             where $\Phi_i$ is the wave function $S_1 S_2 S_3\,\Psi_i$, and $\Phi_i^{(3)}$ the wave function obtained
             from the third diagram in the Faddeev equation. Both of them can be collected beforehand when solving the equation
             and are implemented in the form factor diagram simply through evaluation at the proper incoming momenta, such that
             only the product $S_3 \,\Gamma^\mu$ (which is independent of the loop momentum~$q$) needs to be computed explicitly.

             Again the diagrams for $a=1,2$ can be inferred from the $a=3$ diagram through permutations, namely:
             \begin{equation}\label{current:permutations}
             \begin{split}
                 J^{(1)}_{n'n} &= \left[ \mathcal{M}' J^{(3)} {\mathcal{M}'}^T \right]_{n'n} \left[ \mathcal{M}' \mathsf{F}^{(3)} {\mathcal{M}'}^T \right]_{n'n} \,, \\
                 J^{(2)}_{n'n} &= \left[ \mathcal{M}'' J^{(3)} {\mathcal{M}''}^T \right]_{n'n} \left[ \mathcal{M}'' \mathsf{F}^{(3)} {\mathcal{M}''}^T \right]_{n'n} \,.
             \end{split}
             \end{equation}
             To prove this, one writes down the expressions for $J^{(a=1,2)}$ analogous to Eqs.~\eqref{current:diagram-3} and \eqref{current:diagram-kernel-3}
             with respective momentum dependencies on $p$, $q$, $P$ and $Q$;
             exploits the doublet transformation properties \eqref{perm-tf-doublet} for the amplitudes;
             expresses all internal momenta through $p'$ and $q'$ (for $a=1$) or $p''$ and $q''$ ($a=2$) from Eq.~\eqref{perm-momenta};
             and replaces the $\mathbb{R}^4\times \mathbb{R}^4$ integration over $\{p, q\}$ by integrating
             over $\{p', q'\}$ or $\{p'', q''\}$, respectively.

             Evaluation of Eq.~\eqref{current:permutations} finally yields the following expressions for the total nucleon electromagnetic current of Eq.~\eqref{current:total}:
             \begin{equation} \label{current-proton-neutron}
             \begin{split}
                 &\text{Proton}:    J^\mu_{\delta'\delta} = \left[ \,2 \,J_{11}^{(3),\text{SUM}}\right]^\mu_{\delta'\delta}\,, \\
                 &\text{Neutron}:   J^\mu_{\delta'\delta} = \left[ J_{22}^{(3),\text{SUM}} - J_{11}^{(3),\text{SUM}}\right]^\mu_{\delta'\delta}\,.
             \end{split}
             \end{equation}
             If we denote by $G^{AA}$ the form factors obtained via Eq.~\eqref{obtain-ffs} from the matrix element $J_{11}^{(3)}$ involving the mixed-antisymmetric Dirac amplitudes,
             and by $G^{SS}$ those corresponding to the mixed-symmetric overlap $J_{22}^{(3)}$, one obtains at the relation \eqref{ff-relations-from-aa-ss} for the proton and neutron form factors:
             \begin{equation}
                 G^p = 2 \,G^{AA}\,, \quad G^n = G^{SS} - G^{AA}\,.
             \end{equation}

             To conclude this section we note that we encountered difficulties in the form factor calculation in connection with the angular variable $z_2$ of Eq.\,\eqref{mom-variables}.
             The dependence upon $z_{2i}=\widehat{q}_i\cdot\widehat{P}_i$ and $z_{2f}=\widehat{q}_f\cdot\widehat{P}_f$
             in the incoming and outgoing nucleon amplitudes of Eqs.~\eqref{current:diagram-3} and~\eqref{current:diagram-kernel-3}
             is reconstructed from a Chebyshev expansion in the rest frame as obtained from the Faddeev equation.
             Such a reconstruction is sensitive to the numerical accuracy in the Faddeev amplitude and, especially for $z_2$, leads to convergence problems at larger $Q^2$.
             To some extent this is expected, as $|z_{2i}|, |z_{2f}| < \sqrt{1+\tau}$ in the form-factor integral, where $\tau=Q^2/(4M^2)$, i.e. the domain of the expansion is no longer
             bounded by a unit circle and grows with the photon momentum.
             Eventually, with more powerful computing resources, such issues can be avoided altogether by solving the Faddeev equation in each moving frame anew,
             cf.~Ref.~\cite{Maris:2005tt} in the context of the pion.
             While the dependence on $z_2$ in the rest frame is very weak and could be neglected in the form factor calculation,
             its inclusion is still necessary to avoid issues with charge conservation at $Q^2=0$.
             To account for this we perform the full Chebyshev resummation in $z_{2i}$, $z_{2f}$ at each $Q^2$ only if
             the conditions
             \begin{equation} \label{Chebyshev-treatment}
                  x_i := \frac{p_i^2}{4} + \frac{q_i^2}{3} > Q^2\,, \quad
                  x_f := \frac{p_f^2}{4} + \frac{q_f^2}{3} > Q^2
             \end{equation}
             are satisfied. By virtue of Eq.~\eqref{Chebyshev-treatment}, the full angular dependence is taken into account at $Q^2=0$,
             whereas at large $Q^2$ the photon momentum acts as a cutoff on the Chebyshev expansion
             and only the zeroth moments are retained.
             This procedure resolves the problem mentioned above.

    \subsection{Quark-photon vertex} \label{app:qpv}

            The only ingredient in the form factor diagrams of Fig.~\ref{fig:faddeev-current-specific} which has not already been defined in connection with the Faddeev equation
            is the dressed quark-photon vertex $\Gamma^\mu_\text{q}(k,Q)$.
            Its general expression is derived from the Ward-Takahashi identity
                  \begin{equation}\label{qpv-wti}
                      Q^\mu \,\Gamma^\mu_\text{q}(k,Q) = S^{-1}(k_+)-S^{-1}(k_-)
                  \end{equation}
            and by imposing regularity at $Q^2=0$. It is expressed by a sum of the Ball-Chiu term~\cite{Ball:1980ay} and a purely transverse contribution:
                  \begin{equation}\label{vertex:BC}
                      \Gamma^\mu_\text{q}(k,Q) =   i\gamma^\mu\,\Sigma_A + 2 k^\mu (i\Slash{k}\, \Delta_A  + \Delta_B) + \Gamma^{\mu}_T\,,
                  \end{equation}
            where $k_\pm = k \pm Q/2$ are the incoming and outgoing quark momenta and
            $A(k^2)$ and $B(k^2)=M(k^2)A(k^2)$ are the dressing functions of the inverse quark propagator
            $S^{-1}(k)=i \Slash{k}\,A(k^2) + B(k^2)$, with
                 \begin{equation*}\label{QPV:sigma,delta}
                     \Sigma_F := \frac{F(k_+^2)+F(k_-^2)}{2} , \quad  \Delta_F := \frac{F(k_+^2)-F(k_-^2)}{k_+^2-k_-^2}.
                  \end{equation*}
             The transverse part can be written as
             \begin{equation}\label{transverse-vertex}
             \begin{split}
                  -i\Gamma^\mu_T &=    \gamma^\mu_T\,\big( f_1  + if_2\, \Slash{Q} \big) + \\
                                  & + if_3\, k\!\cdot\! Q\,\textstyle\frac{1}{2}  \left[\gamma^\mu_T, \,\Slash{k} \right]      +  f_4\,\textstyle\frac{1}{2}  \left[\gamma^\mu_T, \,\Slash{k}_T \right] \Slash{Q} \,  + \\
                                  & + k^\mu_T \,\big( i f_5 + f_6\,k\! \cdot\! Q\,\Slash{Q}  + f_7\,\Slash{k}  + i f_8 \,\Slash{k}_T \, \Slash{Q} \big)\,,
             \end{split}
             \end{equation}
             where the $f_i(k^2, \,k\cdot Q, \,Q^2)$ are scalar dressing functions and
             $\gamma^\mu_T$, $k^\mu_T$ are transverse with respect to the photon momentum $Q$.
             While this basis decomposition is not orthogonal, it provides a simple representation for the Ball-Chiu vertex if
             it is used to describe the vertex in total.

             The quark-photon vertex is obtained self-consistently from its inhomogeneous Bethe-Salpeter equation~\cite{Maris:1999bh}, given by
             \begin{equation}\label{qpv-bse}
             \begin{split}
                  \Big[&\Gamma^\mu_\text{q}(k,Q)-Z_2\,i\gamma^\mu\Big]_{\alpha\beta}  = \\
                  & = \frac{4}{3} \int\limits_{k'} K_{\alpha\alpha'\beta'\beta}\,
                   \left[ S(k_+') \,\Gamma^\mu_\text{q}(k',Q)\,S(k_-')\right]_{\alpha'\beta'}\,,
             \end{split}
             \end{equation}
             where $Z_2$ is the quark renormalization constant and $K_{\alpha\alpha'\beta'\beta}$ the rainbow-ladder kernel of Eq.~\eqref{RLkernel}
             that also appears in the quark DSE and the covariant Faddeev equation.

             To solve the inhomogeneous BSE numerically it is convenient to express the vertex in an orthonormal basis, for instance
             \begin{equation}
                \tau_i^\mu(r,\hat{Q}) \in \left\{ \textstyle{\frac{1}{\sqrt{2}}}\,\gamma^\mu_t, \, r^\mu, \, \hat{Q}^\mu \right\} \times
               \left\{ \mathds{1}\,,\, \Slash{\hat{Q}}\,,\, \Slash{r}\,,\, \Slash{r}\,\Slash{\hat{Q}} \right\}
             \end{equation}
             where, in analogy to the nucleon amplitude decomposition of Eq.~\eqref{amplitude:reconstruction}, we defined a relative momentum transverse to the photon momentum
             by the unit vector $r :=\widehat{k_T}$, and in addition a $\gamma-$matrix which is transverse to both $Q$ and $r$, i.e. $\gamma^\mu_t := \gamma^\mu_T - r^\mu \Slash{r}$.
             The basis elements satisfy the orthogonality relation
             \begin{equation}
                 \textstyle{\frac{1}{4}}\, \text{Tr} \,\left\{ \conjg{\tau}_i^\mu \, \tau_j^\mu  \right\} = \delta_{ij}\,.
             \end{equation}
             Upon projecting the kernel onto this basis it becomes clear that the equations for the
             transverse $(\sim \gamma^\mu_t, r^\mu)$ and longitudinal 
             parts $(\sim \hat{Q}^\mu)$ decouple and can be solved independently.
             The longitudinal result reproduces the longitudinal projection
             of the Ball-Chiu vertex and thereby the Ward-Takahashi identity,
             and purely longitudinal terms do not contribute to nucleon form factors because of current conservation: $Q^\mu J^\mu = 0$.
             Hence it is sufficient to consider the eight transverse elements alone.
             The inhomogeneous BSE self-consistently generates a timelike vector-meson pole in the quark-photon vertex
             at $Q^2 =-m_\rho^2$~\cite{Maris:1999bh}.

\end{appendix}


\bigskip

\bibliographystyle{apsrev-mod}

\bibliography{lit-nucleon}

\end{document}